\documentclass[12pt,letterpaper]{article}
\usepackage{color}
\usepackage{latexsym}
\usepackage{amsthm}
\usepackage{amsmath}
\usepackage{amssymb}
\usepackage{graphicx,array}
\usepackage{hyperref}

\def\simgt{\mathrel{\lower2.5pt\vbox{\lineskip=0pt\baselineskip=0pt
           \hbox{$>$}\hbox{$\sim$}}}}
\def\simlt{\mathrel{\lower2.5pt\vbox{\lineskip=0pt\baselineskip=0pt
           \hbox{$<$}\hbox{$\sim$}}}}

\newcommand{\be}{\begin{equation}}
\newcommand{\ee}{\end{equation}}
\newcommand{\bea}{\begin{eqnarray}}
\newcommand{\eea}{\end{eqnarray}}
\newcommand{\Eq}[1]{Eq.~(\ref{#1})}

\newcommand{\Sec}[1]{Sec.~\ref{#1}}

\newcommand{\Fig}[1]{Fig.~(\ref{#1})}
\newcommand{\App}[1]{App.~\ref{#1}}

\newenvironment{changemargin}[2]{%
  \begin{list}{}{%
    \setlength{\topsep}{0pt}%
    \setlength{\leftmargin}{#1}%
    \setlength{\rightmargin}{#2}%
    \setlength{\listparindent}{\parindent}%
    \setlength{\itemindent}{\parindent}%
    \setlength{\parsep}{\parskip}%
  }%
  \item[]}{\end{list}}

\definecolor{nicered}{rgb}{0.7,0.1,0.1}
\definecolor{nicegreen}{rgb}{0.1,0.5,0.1}
\hypersetup{colorlinks,citecolor= nicegreen,linkcolor= nicered}

\begin{document}
\hfill

\vspace{4cm}

\begin{center}
{\LARGE\bf
Higgs and Dark Matter Hints
\\[3mm]
of an Oasis in the Desert
}\\
\bigskip\vspace{1cm}{
{\large Clifford Cheung$^{1,2}$, Michele Papucci$^{1,2}$, Kathryn M. Zurek$^{3}$}
} \\[7mm]
 {$^1$\it Department of Physics, University of California,
          Berkeley, CA 94720} \\
 {$^2$\it Theoretical Physics Group, Lawrence Berkeley National Laboratory,
          Berkeley, CA 94720} \\
 {$^3$\it Michigan Center for Theoretical Physics, University of Michigan, Ann Arbor, MI 48109}
 \end{center}
\bigskip
\centerline{\large\bf Abstract}

\begin{quote} \small

Recent LHC results suggest a standard model (SM)-like Higgs boson in the
vicinity of 125 GeV with no clear indications yet of physics beyond the SM.  At the same time, the SM is incomplete, since additional
dynamics are required to accommodate cosmological dark matter (DM).
In this paper we show that interactions between weak scale DM and the
Higgs which are strong enough to yield a thermal relic abundance consistent with observation can
easily destabilize the electroweak
vacuum or drive the theory into a non-perturbative regime at a low scale.
As a consequence, new physics---beyond the DM itself---must enter
at a cutoff well below the Planck scale and in some cases as low as ${\cal O}$(10 - 1000 TeV), a range relevant to indirect probes of flavor and CP violation. In addition, this cutoff is correlated with the DM mass and scattering cross-section in a parameter space which will be probed experimentally in the near term.
Specifically, we consider the SM plus additional spin 0 or 1/2 states
with singlet, triplet, or doublet electroweak quantum numbers and quartic or Yukawa couplings to the Higgs boson.  We derive explicit expressions for the full two-loop RGEs and
one-loop threshold corrections for these theories.

\end{quote}

\newpage

\section{Introduction}

With the quest for the elusive Higgs boson approaching a conclusive end, it is crucial that we evaluate the implications of its imminent discovery or exclusion for beyond the standard model (SM) physics.  In the past, a conservative approach to this question has been to assess the consistency of the SM assuming a vast desert above the weak scale.  Within this framework, numerous authors~\cite{massgrave1,Ellis:2009tp,EliasMiro:2011aa,Giudice:2011cg} have analyzed the SM with respect to the stability of the electroweak symmetry breaking vacuum and the perturbativity of the underlying dynamics.  As is well known, the relevant physics is determined by running couplings into the ultraviolet using renormalization group equations (RGEs).  Hence, results depend sensitively on the weak scale boundary conditions, among which the mass of the Higgs boson is perhaps most critical.  Recent experimental results from ATLAS~\cite{ATLAS} and CMS~\cite{CMS} suggest a value of the Higgs mass around 125 - 126 GeV. This value  corresponds to the SM with perturbative couplings up to the Planck scale, and a metastable electroweak vacuum with a lifetime longer than the age of the Universe~\cite{Ellis:2009tp,EliasMiro:2011aa}.  

Despite its successes, however, the SM is almost certainly incomplete, since new physics is required below the Planck scale in order to accommodate observed phenomena such as dark matter, baryogenesis, neutrino masses, and the QCD theta parameter.  In principle, the new dynamics may couple directly to the Higgs boson, therefore inducing  deviations from the usual vacuum stability and perturbativity bounds of the SM.  At the same time, many of these theories, \emph{e.g.} the neutrino seesaw and the QCD axion, are accommodated by rather high scale dynamics, in which case there will be a minimal effect on the running of couplings.  A notable exception to this is weakly interacting massive particle DM, whose mass scale is constrained to be low by thermal freeze-out.

In this paper we carry out a general analysis of vacuum stability and perturbativity in the SM augmented by weak scale DM.  We consider additional spin 0 or 1/2 states with singlet, triplet, or doublet electroweak quantum numbers, and including all quartic and Yukawa couplings to the Higgs boson consistent with the stability of DM.  For our analysis we have derived two-loop RGEs and one-loop threshold corrections for these theories.  We find that fermionic DM tends to destabilize the electroweak vacuum, such that new dynamics below the Planck scale is required ensure stability.  On the other hand, scalar DM tends to stabilize the vacuum, though demanding perturbativity by itself may require new physics at intermediate scales.  Moreover, imposing a thermal relic abundance fixes a minimum value for the new quartic or Yukawa couplings which can have a substantial effect on vacuum stability and perturbativity.  This effect is correlated with the DM mass and scattering cross-section in a parameter space which will be accessed by DM direct detection experiments in the near future.  In many cases the required cutoffs for these theories can be as low as 10 - 1000 TeV, a range of energy scales which is reachable with indirect searches, such as those probing flavor and CP violating interactions.  Our results are consistent with previous analyses of specific scalar~\cite{RamseyMusolf,FileviezPerez} and fermionic~\cite{Chen} DM models.

 \begin{figure}[t]
\label{fig:SM}
\begin{center} 
\includegraphics[scale=1.3]{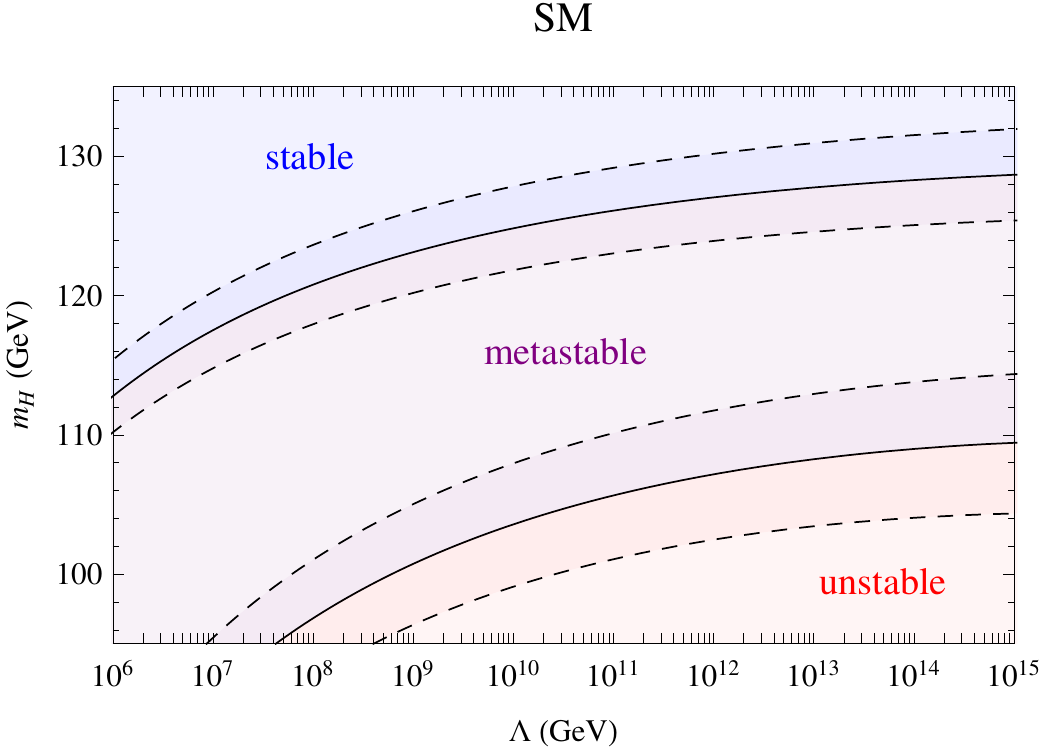}
\end{center}
\caption{Vacuum structure of the SM as a function of the Higgs boson mass.  Regions of stability/metastability/instability are denoted in blue/purple/red respectively. The solid lines indicate central values while the dotted lines indicate $\pm 2 \sigma$ error bars on the experimental measurement of the top quark mass. }
\end{figure}

Let us elaborate briefly on the notion of vacuum stability in the SM and beyond.  Depending on the Higgs quartic $\lambda_H$ at the scale $\Lambda$, our vacuum may be any of the following:
\begin{itemize}
\item[i)] Stable ($\lambda_H > 0$).  The vacuum is the absolute minimum and will never decay.
\item[ii)] Metastable ($0 > \lambda_H > \hat\lambda_H$).  The vacuum is not the absolute minimum, but its lifetime is longer than the age of the Universe.
\item[iii)] Unstable ($\hat\lambda_H > \lambda_H$).  The vacuum is not the absolute minimum, and it decays within the age of the Universe.  
\end{itemize}
Here the critical coupling $\hat\lambda_H$ is determined by the requirement that the tunneling rate per unit volume is comparable to the age of the Universe.  In particular, we demand that $H^4=\Gamma$, where $H^{-1} \simeq 3.7$ Gyr and $\Gamma$ reads,
\bea
\Gamma &=&  {\rm max}\left[  R^{-4} \exp(-16\pi^2/ 3 |\hat\lambda_H|) \right] \bigg|_{R^{-1}< \Lambda}.
\label{eq:stabtypes}
\eea
Here $R$ is the characteristic length scale of the bounce, which is bounded by the cutoff.  As we will elaborate on later, the vacuum structure may be more complicated if the new physics includes additional scalar particles.

It is possible that our vacuum resides in a stable or metastable regime, but the unstable regime  is of course excluded by our existence.  In much of our analyses, it will be convenient to summarize the nature of the vacuum by depicting the ``metastability band'' in parameter space defined by the region 
\bea 
\hat \lambda_H<\lambda_H < 0.
\eea
Note that for theories in which supersymmetric dynamics enters at the cutoff $\Lambda$, \emph{i.e.}~split or high-scale supersymmetry, absolute stability is required by the fact that the potential is positive semi-definite in all field directions, so $\lambda_H \geq0$.   

By running the RGEs into the ultraviolet, subject to infrared boundary conditions, it is possible to determine the scale $\Lambda$ at which the quartic coupling crosses through the  metastability band.  For example, in \Fig{fig:SM} we plot the scale $\Lambda$ indicating the onset of stability/metastability/instabilility in the SM as a function of the $m_H$.  Our results are in nice agreement with existing calculations in the literature~\cite{Ellis:2009tp,EliasMiro:2011aa,Giudice:2011cg}.
 
The remainder of this paper is as follows.  In \Sec{sec:models} we briefly summarize the set of models to be studied and establish notational conventions.  We present our methodology in \Sec{sec:methodology} regarding the running of couplings and evaluation of DM properties.  Finally, in \Sec{sec:results} we discuss our results.  The two-loop RGEs and one-loop threshold corrections for these theories are presented in \App{app:RGEs} and \App{app:thresholds}.

 \begin{figure}[t]
\includegraphics[scale=1]{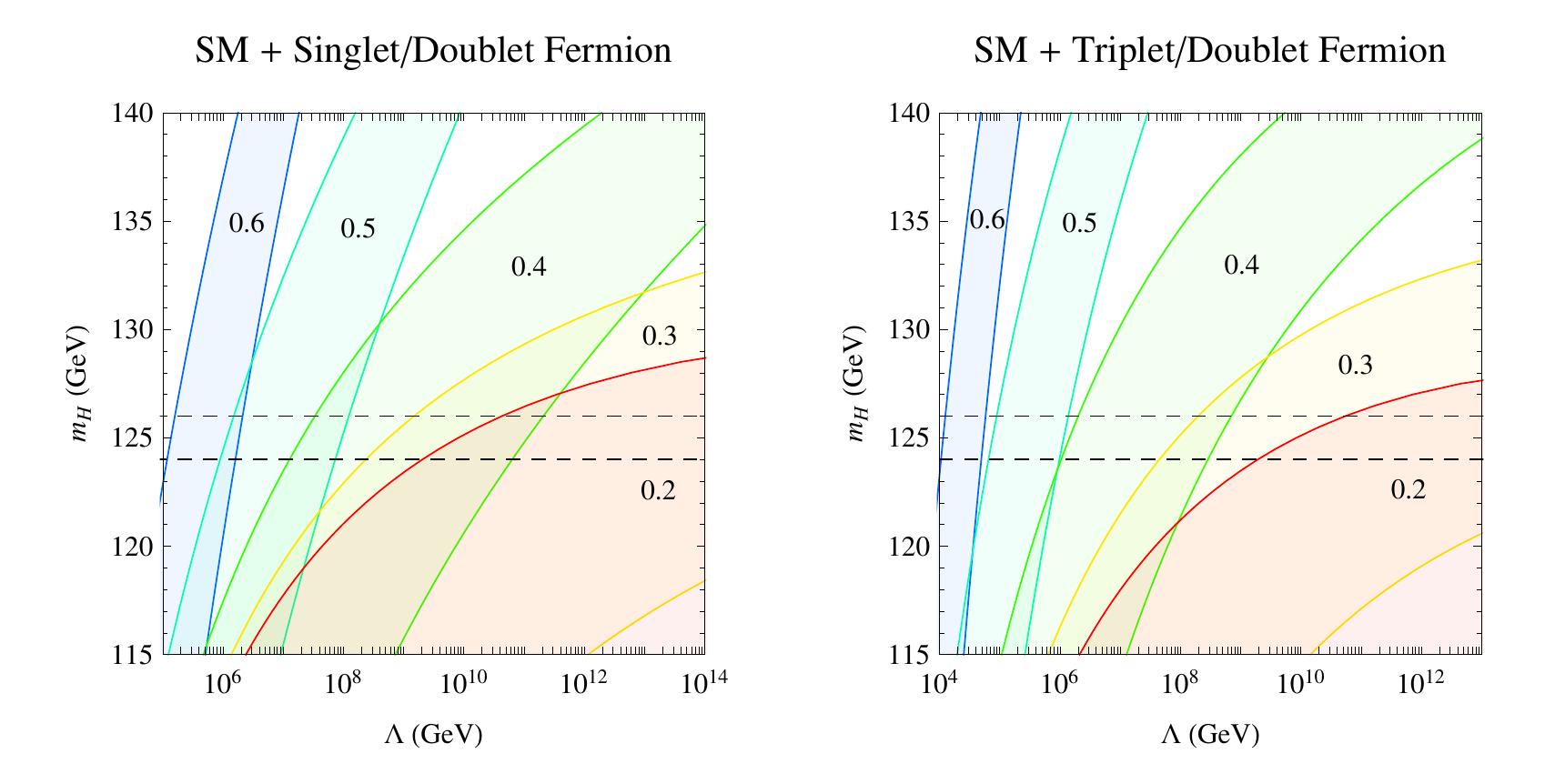}
\caption{Metastability bands for SM + singlet/doublet and triplet/doublet fermion, shown as a function of the Higgs mass.  Regions above/below each band are stable/unstable.
Each band is labeled with the corresponding value of the Yukawa coupling, $y_{S,T} = y_{S,T}^c$.  
The dashed lines correspond to the value of $m_H$ suggested by~\cite{ATLAS,CMS}.\label{fig:STD_mh_fermion}}
\end{figure}

\begin{figure}[t]
\includegraphics[scale=1]{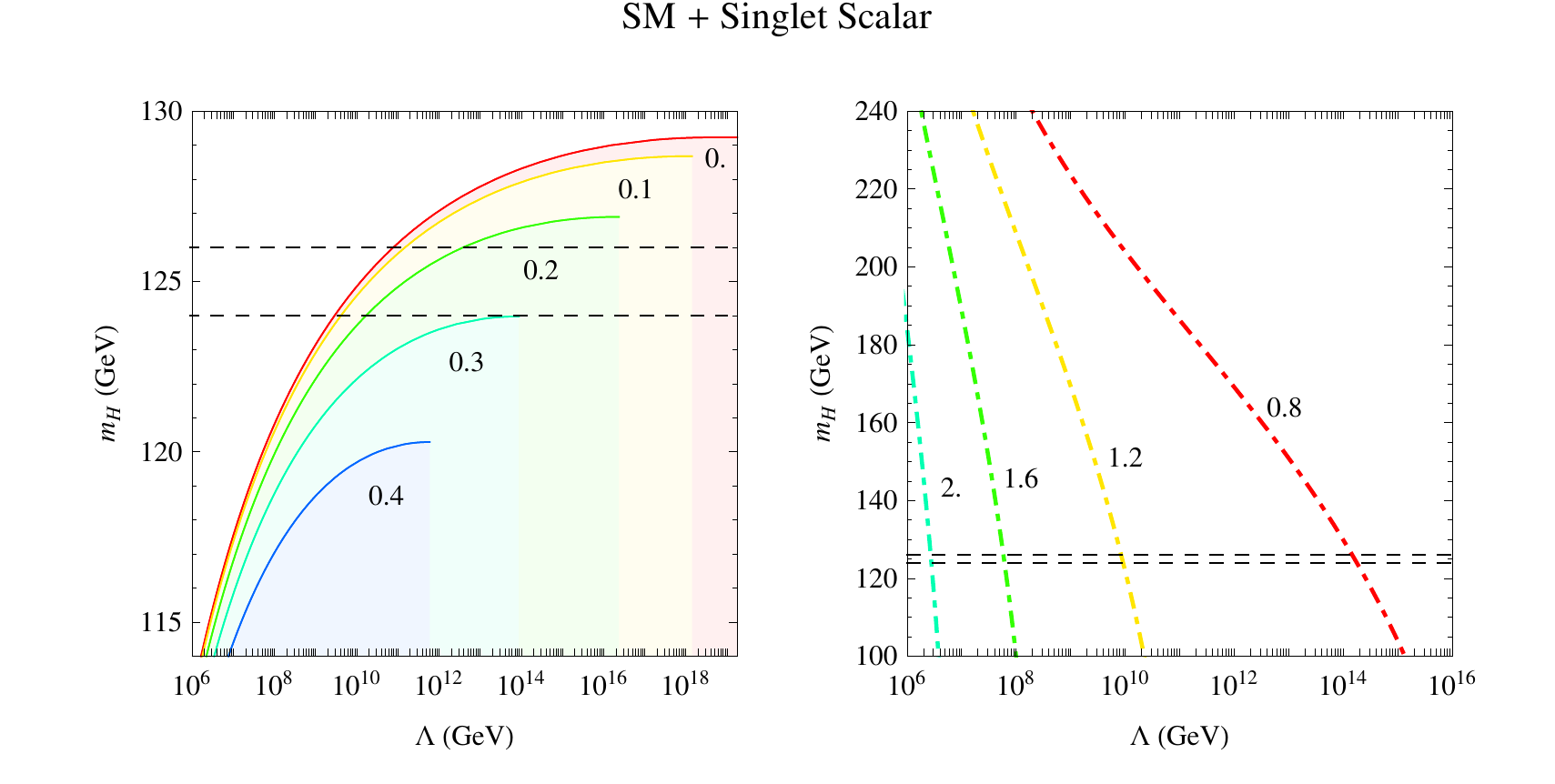}
\caption{Higgs mass bounds as a function of the cutoff, for the SM + singlet scalar. The left panel depicts metastability bands and the right panel depicts the scale at which the largest coupling in the theory becomes non-perturbative.  Each label denotes the corresponding value of the cross-quartic coupling, $\kappa_S$, and we have fixed the self-quartic coupling, $\lambda_S=0$.
\label{fig:S_mh_scalar}}
\end{figure}

\begin{figure}[t]
\includegraphics[scale=1]{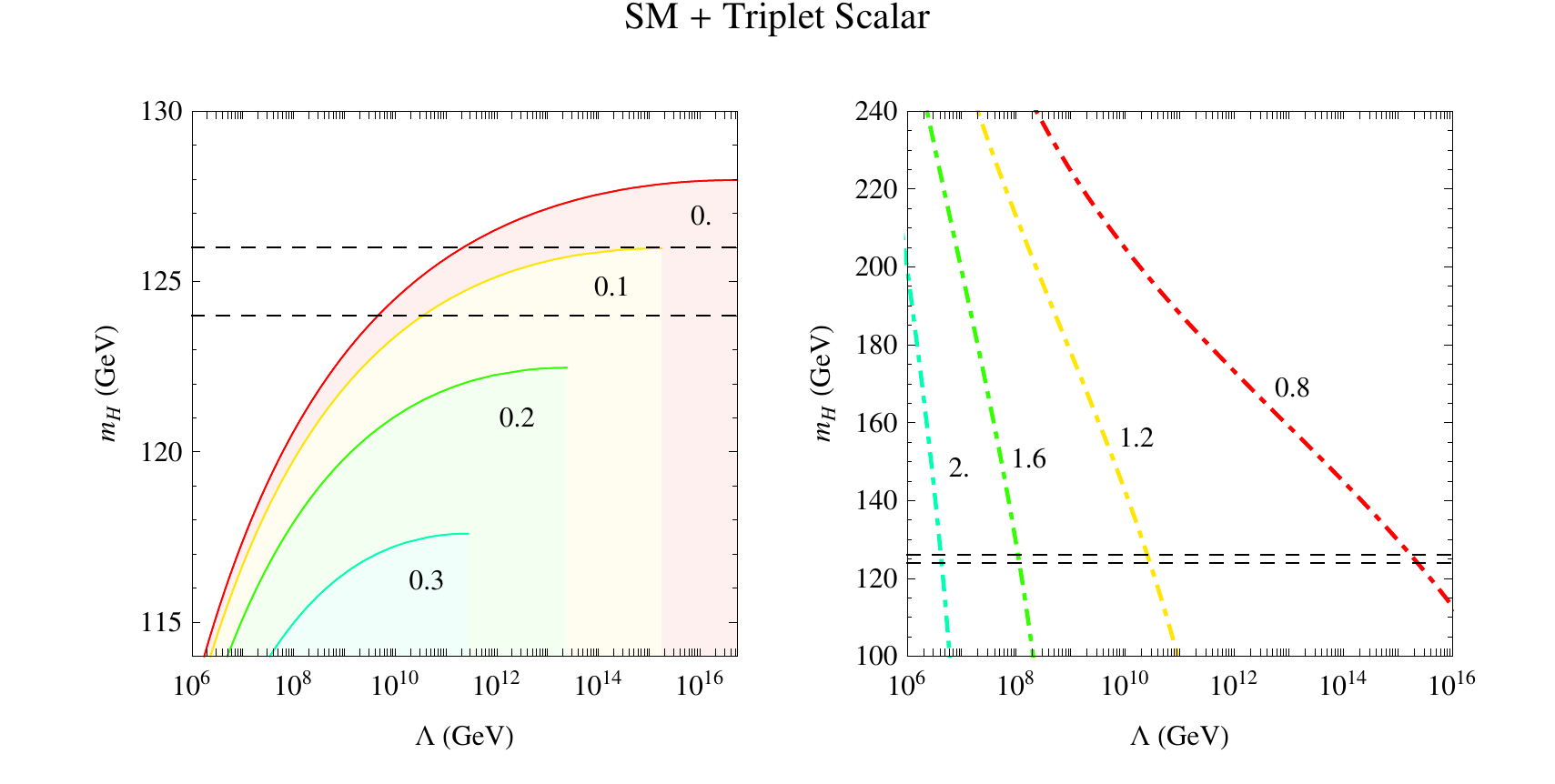}
\caption{Same as \Fig{fig:S_mh_scalar} but for SM + triplet scalar and varying $\kappa_T$ with $\lambda_T$=0.
\label{fig:T_mh_scalar}}
\end{figure}

\begin{figure}[t]
\includegraphics[scale=1]{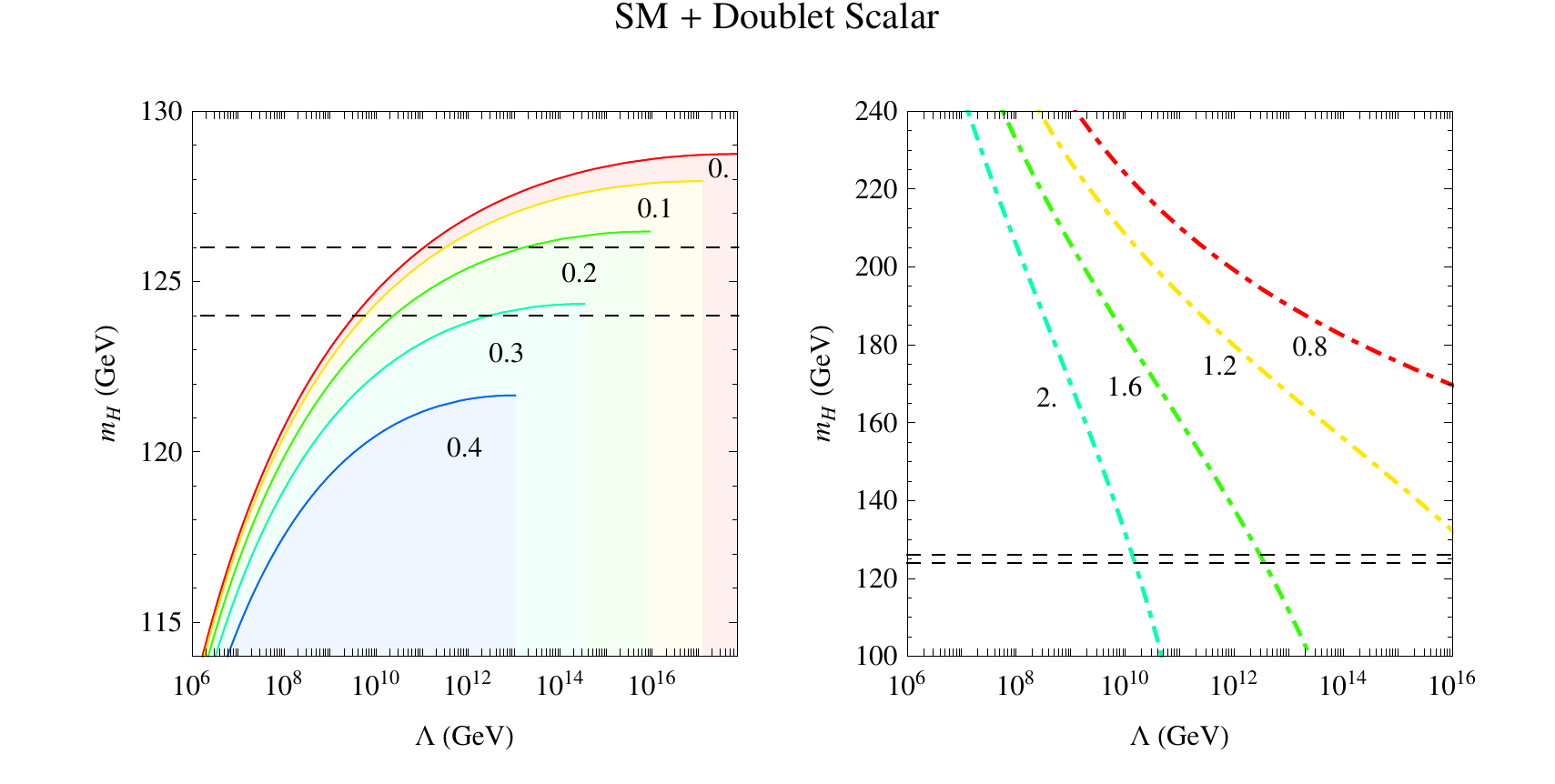}
\caption{Same as \Fig{fig:S_mh_scalar} but for SM + doublet scalar and varying $\kappa_D$ with $\lambda_D = \lambda_{D}' = \kappa_{D}'$=0.
\label{fig:D_mh_scalar}}
\end{figure}

\section{Models}
\label{sec:models}

In this section we briefly summarize the models we will analyze and provide notational conventions.  We choose the following normalization for the SM quartic and Yukawa couplings,
\bea
-{\cal L} &=& y_t q \epsilon H t^c + y_b q H^\dagger b^c + y_\tau \ell H^\dagger \tau^c +  \frac{\lambda_H}{2} (|H|^2 - v^2)^2,
\label{eq:SMnotation}
\eea
where $v=174.1$ GeV is the Higgs vacuum expectation value.  Throughout, we ignore the effects of the light fermion generations because they are negligible for our results.

We will augment the SM with new states, sending ${\cal L} \rightarrow {\cal L} + \Delta{\cal L}$.  
Throughout, if a particle $X$ is a fermion, then $y_X$ will denote its Yukawa coupling of $X$ to the Higgs.  If a particle $X$ is a scalar, then $\lambda_X$ will denote its self-quartic coupling, while $\kappa_X$ will denote its cross-quartic coupling to the Higgs.  Finally, the mass parameter for the particle $X$ will be denoted by $m_X$.

For the case of new fermions we consider the following theories:
\bea
\nonumber\\
\textrm{singlet/doublet fermion:} \quad
-\Delta {\cal L} &=& \frac{1}{2}m_S S^2 + m_D D D^c +y_S H S D + y_S^c H^c S D^c \nonumber\\
\nonumber\\ 
\textrm{triplet/doublet fermion:} \quad
-\Delta {\cal L} &=& \frac{1}{2}m_T T^2 + m_D D D^c +y_T H T D + y_T^c H^c T D^c,\nonumber \\
\eea
where we have defined $H^c \equiv \epsilon H^*$, and we have included all renormalizable operators permitted by gauge symmetry.  Here $S$ and $T$ are Majorana fermions and $D$ and $D^c$ form a Dirac fermion.  As a consequence of the additional Yukawa couplings to the Higgs, the new fermionic states will generally mix after electroweak symmetry breaking, although there is a preserved $\mathbb{Z}_2$ symmetry which makes the lightest among these fields a DM candidate.
We do not consider the fully mixed singlet/triplet/doublet theory, since all the qualitative features of this model are already evident in the singlet/doublet and triplet/doublet cases, while the proliferation of parameters would make results difficult to present.
 
The fermionic theories above are of course a generalization of the bino, wino, and Higgsino sector of the MSSM.  
In the chosen parameterization, the precise correspondence between these theories and the MSSM is\bea
\begin{array}{ccc}
S &\leftrightarrow & \tilde B \\
T &\leftrightarrow & \epsilon \tilde W \epsilon^T \\
D &\leftrightarrow & \epsilon \tilde H_d \\
D^c &\leftrightarrow & \epsilon \tilde H_u 
\end{array}
\qquad
\begin{array}{ccc}
m_S &\leftrightarrow& M_1 \\
m_T &\leftrightarrow& M_2 \\
m_D  &\leftrightarrow&\mu
\end{array}
\qquad
\begin{array}{ccc}
y_S &\leftrightarrow&  \tilde{g}'_d / \sqrt{2}\\
y_S^c &\leftrightarrow&  \tilde{g}'_u / \sqrt{2}\\
y_T &\leftrightarrow&  -\tilde{g}_d / \sqrt{2}\\
y_T^c &\leftrightarrow&  \tilde{g}_u / \sqrt{2}\\
\end{array},
\eea
using the notation of~\cite{Giudice:2004tc}.  Furthermore, in the limit of exact supersymmetry, $\tilde{g}^{(\prime)}_u = g^{(\prime)} \sin\beta$ and $\tilde{g}^{(\prime)}_d = g^{(\prime)} \cos\beta$.

For the case of new scalars we consider the following theories:
\bea
\nonumber\\
\textrm{singlet scalar:} \quad
-\Delta {\cal L} &=& \frac{1}{2}m_S S^2 + \frac{\lambda_S}{2} S^4 + \frac{\kappa_S}{2} S^2 |H|^2 \nonumber\\
\nonumber\\
\textrm{triplet scalar:} \quad
-\Delta {\cal L} &=& \frac{1}{2}m_T T^2 + \frac{\lambda_T}{2} T^4 + \frac{\kappa_T}{2}  T^2 |H|^2 \nonumber\\
\nonumber\\
\textrm{doublet scalar:} \quad
-\Delta {\cal L} &=& m_D |D|^2 + \frac{\lambda_D}{2} |D|^4 + \frac{\kappa_D}{2} |D|^2|H|^2 + \frac{\kappa_D'}{2} |D H^\dagger|^2.\nonumber\\
\eea

Here $S$ and $T$ are real scalars while $D$ is a complex scalar.  Contrary to the fermionic case, the pure doublet scalar case can have direct couplings to the Higgs and therefore can be considered alone without including singlets or triplets.
For the singlet and triplet theories we have included all operators permitted by gauge symmetry subject to a discrete $\mathbb{Z}_2$ symmetry under which $S$ and $T$ are odd.   
For the doublet scalar theory we have assumed a Peccei-Quinn symmetry which acts oppositely on $D$ and $H$ in order to reduce the number of possible operators to a manageable number.  Hence, all of these theories carry a $\mathbb{Z}_2$ symmetry under which the lightest odd particle is a prospective DM particle.

\section{Methodology}
\label{sec:methodology}

To study vacuum stability and perturbativity in the presence of additional dynamics we have produced code which inputs an arbitrary Lagrangian and outputs the corresponding two-loop RGEs and one-loop weak scale threshold corrections to Higgs and top couplings.
For the two-loop RGEs we employed the $\overline{\rm MS}$ expressions of~\cite{Machacek}, including the corrections discussed in~\cite{Luo}.  We have also computed one-loop weak scale threshold effects for the theories under consideration using the methodology of~\cite{Sirlin}.  All of these results, including the general formulae for the threshold corrections using the notation of~\cite{Luo} and~\cite{Pierce}, are presented in detail in \App{app:RGEs} and \App{app:thresholds}.

As is well-known, the running of $\lambda_H$ is highly sensitive to the weak scale values of $\lambda_H$, $y_t$ and $\alpha_s$.  This is because Higgs boson/top quark loops are the primary contribution driving $\lambda_H$ to be positive/negative in the ultraviolet.  In turn, the strong interactions feed into the top Yukawa coupling.  In our analysis we take the current values of the top quark pole mass and the $\overline{\rm MS}$ running QCD coupling measured at the $Z$ pole~\cite{PDG}:
\bea
m_t &=& 173.2 \pm  0.9 \textrm{ GeV} \; (1\sigma)\\ \nonumber
\alpha_s(m_Z)&=& 0.1184 \pm 0.0007 \; (1 \sigma).
\label{mt_alphas}
\eea
On the other hand the Higgs boson has not been found and its mass has yet to be measured. Nevertheless, we make use of recent experimental results from the LHC. Since almost all of the theoretically allowed region for $m_H$ is excluded at 95\% CL by both ATLAS and CMS, and both collaborations observe an excess at 125 - 126 GeV, assuming that the present excess corresponds to the Higgs boson, we will use
\bea 
m_H &=& 125 \textrm{ GeV},
\label{mH}
\eea
in our analysis.
We relate the top quark and Higgs boson pole masses to $y_t(Q)$ and $\lambda_H(Q)$,  the Yukawa and quartic couplings renormalized at an $\overline{\rm MS}$ scale $Q$, via
\bea
m_t &=& y_t(Q) v  [1+ \delta_t (Q)]^{-1} \\
m_H^2&=& 2 \lambda(Q) v^2 [1+ \delta_H(Q)].
\label{tH_thresholds}
\eea
Here $\delta_{t,H}$ includes one-loop threshold corrections from the SM and new physics contributions.  We present general formulae for the one-loop threshold corrections in \App{app:thresholds}. Furthermore $\delta_{t}$ includes the shift between pole and running mass due to QCD interactions which is known to three loops~\cite{Chetyrkin:1999qi}.

We have chosen our $\overline{\rm MS}$ matching scale $Q$ to be the top pole mass $m_t$.  Our boundary conditions for SM parameters are determined by \Eq{mt_alphas}, \Eq{mH}, and \Eq{tH_thresholds}, where we have run $\alpha_s(m_Z)$ from $m_Z$ to $m_t$ using the SM two-loop RGEs enhanced by the three-loop QCD beta function~\cite{PDG}.  We then run the RGEs up to a scale $Q = \Lambda$ and determine the value of $\lambda_H$ at that scale.  As defined in \Eq{eq:stabtypes}, if $\hat \lambda_H <\lambda_H <0$ then the vacuum is metastable but long-lived, while if $\lambda_H <\hat \lambda_H $ then the vacuum is unstable and decays within the age of the Universe.  

Note that $\lambda_H \geq 0$ is a necessary but not sufficient condition for absolute stability of the vacuum.  
Likewise, $0> \lambda_{H}> \hat \lambda_{H}$ does not guarantee that the lifetime of the vacuum exceeds the age of the Universe.  
The reason for this is that in theories with new scalar fields, the vacuum structure is enriched.   Absolute stability of the vacuum implies further conditions on scalar field theories: the self-quartic couplings satisfy $\lambda_S ,\lambda_T,\lambda_D \geq 0$, while the cross-quartic couplings are bounded by
\bea
\kappa_S &\geq& -2\sqrt{\lambda_H \lambda_S}\\
\kappa_T &\geq& -2\sqrt{\lambda_H \lambda_T}\\
\kappa_D +\kappa_D' &\geq& -2\sqrt{\lambda_H \lambda_D},
\eea
for all scales below the cutoff $\Lambda$. If these criteria are not satisfied, then there will exist field directions in which the potential is unbounded from below at large field values.  Alternatively, if the additional scalar fields acquire vacuum expectation values, this can also substantially alter the stability of the vacuum \cite{Strumia2012}.   However, we will not consider this possibility since our interest is in DM.

In addition to the question of stability we also investigate the perturbativity of interactions in the ultraviolet.  For our purposes we define perturbativity to be the criterion that for each coupling $g$, the contribution of $g$ to its own beta function is bounded by unity.  Precisely, we require that $d g /d\log Q = \beta(g) < 1$, which can be read off trivially from RGEs presented in \App{app:RGEs}.  As we will see, the constraint of perturbativity will be largely unimportant for the case of new fermionic states, but can play an essential role in determining the cutoff for theories with new scalars.   For the scalars, the perturbativity bounds on the scalar couplings are
\bea
\kappa_S^2  <  16\pi^2 /4, &&\quad  \lambda_S^2 < 16\pi^2 / 36 \\
\kappa_T^2  <  16\pi^2 /4 ,&&\quad  \lambda_T^2 < 16\pi^2 / 44 \\
\kappa_D^2,\kappa_D'{}^{2}  <  16\pi^2 /2 , &&\quad   \lambda_D^2 < 16\pi^2 / 12.
\label{scalarpert}
\eea

To evaluate the properties of DM in each of these theories, we have implemented each model in {\tt LanHEP}~\cite{LanHEP} and evaluated relic abundances and direct detection cross-sections in {\tt micrOMEGAs}~\cite{micrOMEGAs}.  For the direct detection cross-sections we have employed the most recent lattice results~\cite{Giedt} for the nuclear form factors,
\bea
\begin{array}{ccc}
f_{Tu}^{(p)} = 0.0280 & f_{Td}^{(p)} = 0.0280 & f_{Ts}^{(p)} = 0.0689.
\end{array}
\eea
Nuclear uncertainties will in general affect these results, particularly for the strange quark contribution. 
For our direct detection bounds, we have taken the current limits derived from the XENON100 experiment~\cite{XENON100}.

Lastly, let us briefly comment on electroweak constraints.  For the range of allowed couplings, the theories described in \Sec{sec:models} typically possess violations of custodial symmetry and induce corrections to the $\rho$ parameter.  However, for all of the theories we consider here, we have checked that these effects are well within experimental bounds. We also have checked the corrections to the $S$ parameters across the whole range of parameter space considered here and found them to be small as well, mostly due to the relatively heavy spectra.

\begin{figure}[htbp]
\hspace*{-1cm}
\includegraphics[scale=1]{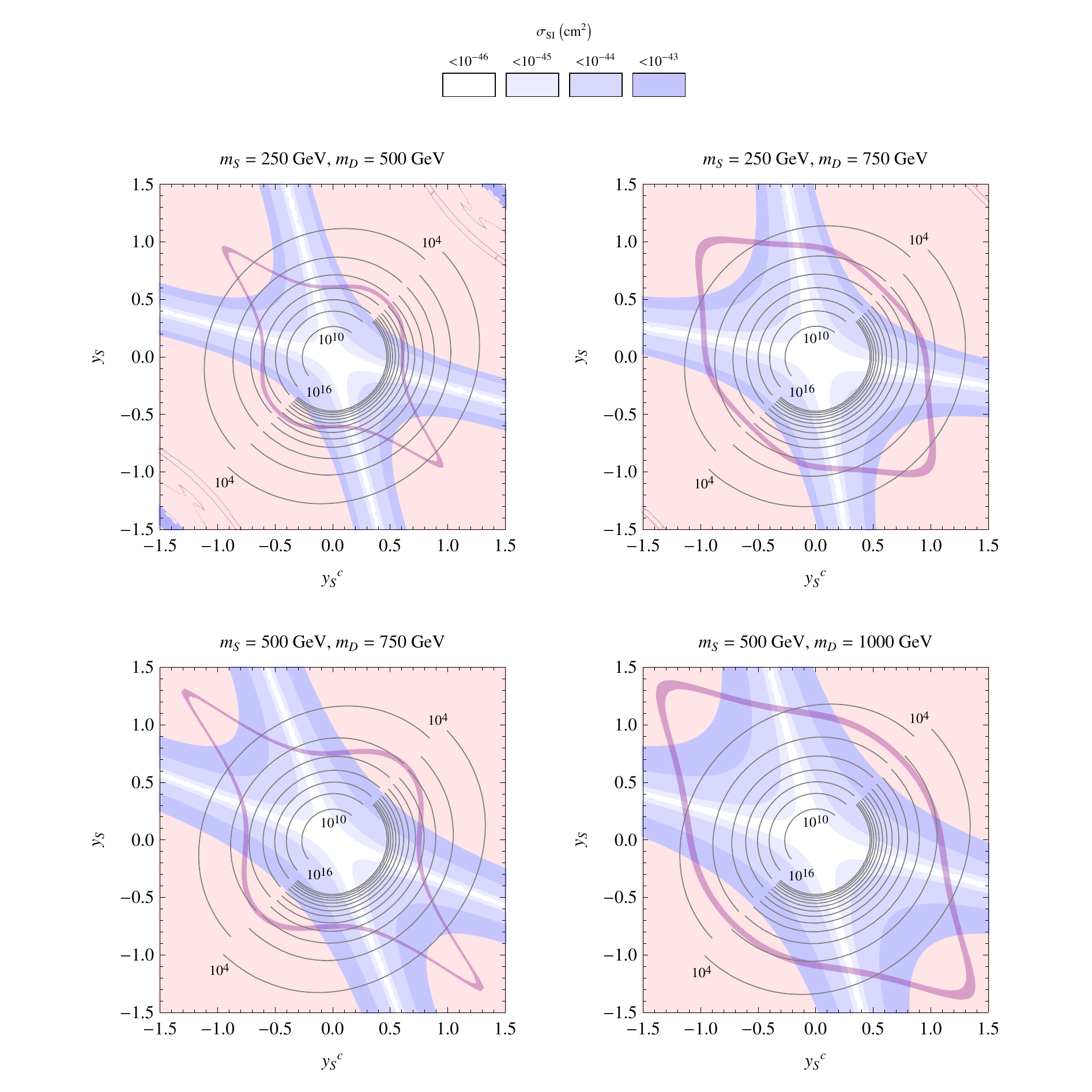}
\caption{Contour plots for SM + singlet/doublet fermion with singlet-like DM, shown in the $(y_S^c, y_S)$ plane at fixed values of $m_S$ and $m_D$.  The purple bands corresponds to $\Omega h^2 = 0.11 \pm 0.01$.  The red/blue regions are excluded/allowed by XENON100, with $\sigma_{SI}$ denoted.  The gray contours in the upper left/lower right quadrants denote the scale $\Lambda$ (GeV) at which the vacuum becomes metastable/unstable.
\label{fig:SD_fermion_S}}
\end{figure}

\begin{figure}[htbp]
\hspace*{-1cm}
\includegraphics[scale=1]{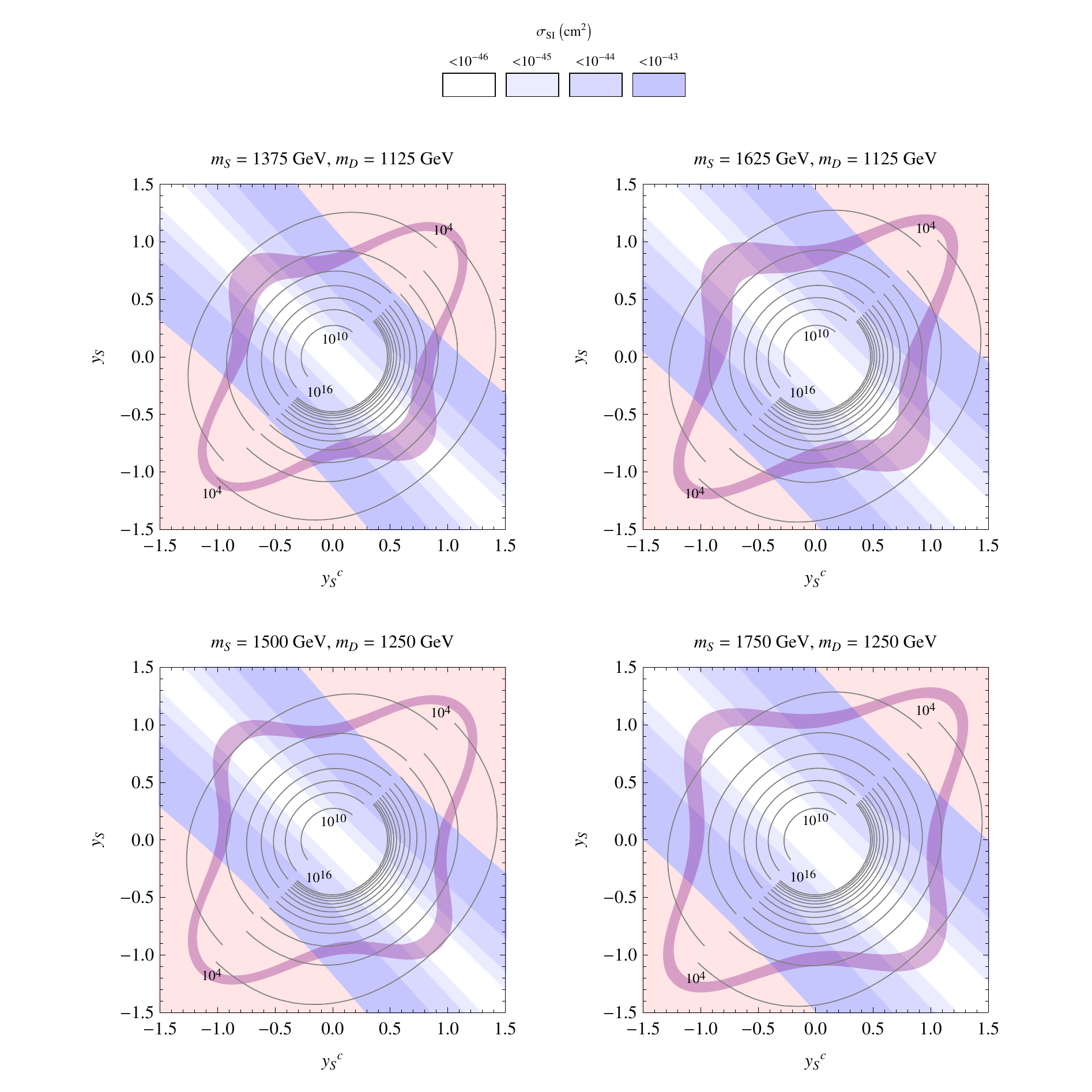}
\caption{Same as \Fig{fig:SD_fermion_S} but for SM + singlet/doublet fermion with doublet-like DM.
\label{fig:SD_fermion_D}}
\end{figure}

\begin{figure}[htbp]
\hspace*{-1cm}
\includegraphics[scale=1]{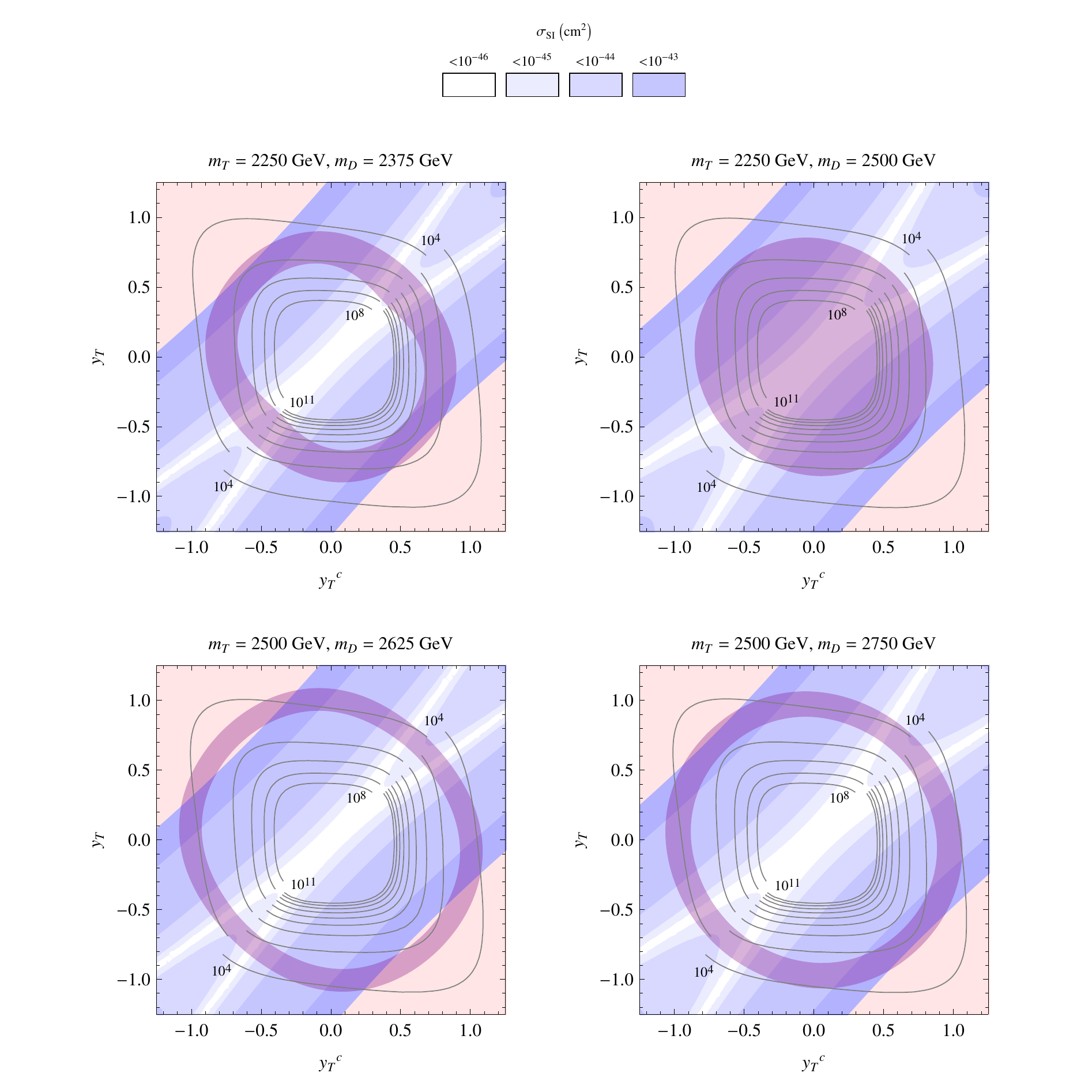}
\caption{Same as \Fig{fig:SD_fermion_S} but for SM + triplet/doublet fermion with triplet-like DM.
\label{fig:TD_fermion_T}}
\end{figure}

\begin{figure}[htbp]
\hspace*{-1cm}
\includegraphics[scale=1]{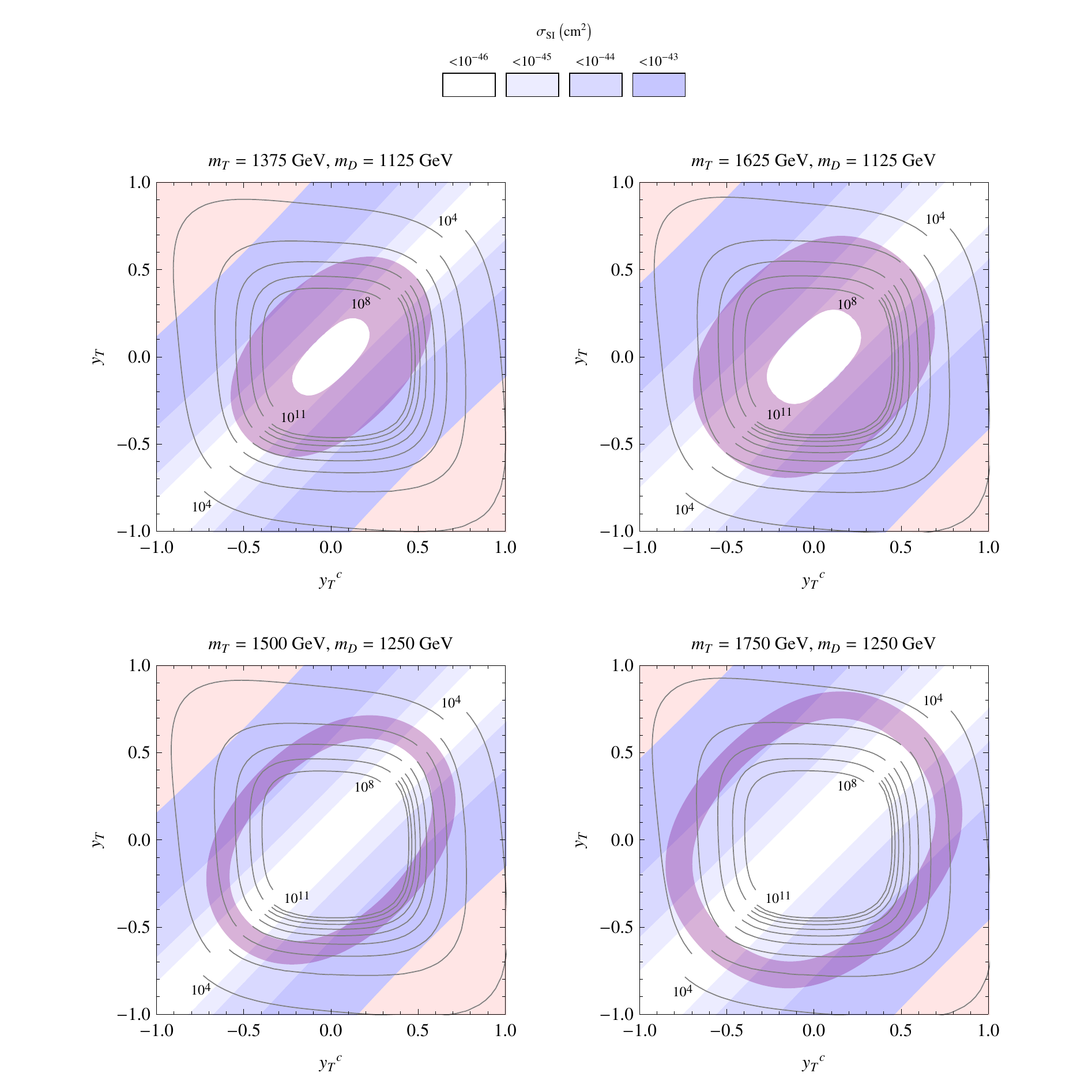}
\caption{Same as \Fig{fig:SD_fermion_S} but for SM + triplet/doublet fermion with doublet-like DM.
\label{fig:TD_fermion_D}}
\end{figure}

\section{Results}
\label{sec:results}

In this section we present the results of our analysis.  We will begin with a general discussion of stability bounds, independent of the fact that the new states may constitute the DM of the Universe.  Afterwards, we will present an analysis dedicated to DM in which we impose considerations from thermal relic abundance and direct detection constraints.

\subsection{Stability, Perturbativity, and New Physics}

We now analyze the effect of new physics on the vacuum as a function of the model parameters input at the weak scale.  Consider \Fig{fig:STD_mh_fermion}, which characterizes the stability of the vacuum in the $(\Lambda, m_H)$ plane for the singlet/doublet and triplet/doublet theories outlined in \Sec{sec:models}.  These plots are similar to the SM plot presented in \Fig{fig:SM}.  Each colored region corresponds to the metastability band, $\hat\lambda_H < \lambda_H < 0$, for a different value of the Yukawa coupling, fixing for simplicity $y_S = y_S^c$ and $y_T = y_T^c$.  Hence, the upper and lower boundary of each band denotes the scale $\Lambda$ at which $\lambda_H =0$ and $\lambda_H = \hat \lambda_H$, respectively.   In this plot we have fixed $m_S = 100$ GeV, $m_T= 200$ GeV, and $m_D = 500$ GeV, entering the one-loop threshold corrections.

 \Fig{fig:STD_mh_fermion} demonstrates how new Yukawa couplings to the Higgs tend to destabilize the vacuum.  Indeed, the scale of instability drops precipitously---to within even a couple of orders of magnitude of the weak scale for $m_H \simeq 125$ GeV---as soon as $y_{S,T}^{(c)} \gtrsim 0.5$.  The reason for this trend is that loops of new fermions tend to drive the Higgs quartic negative in the ultraviolet.  To give a sense of the relative size of these couplings, recall that in the MSSM at $\tan \beta = 1$, the analogous gaugino couplings correspond to $y_S = y_S^c = g'/2 \simeq 0.2$ and $y_T = y_T^c = g/2 \simeq 0.3$.  Lastly, note that the constraint of perturbativity is typically unimportant.  In particular, current Higgs boson exclusions~\cite{ATLAS, CMS} suggest that $m_H$ is in the light range, so $\lambda_H$ is small and remains perturbative.  Likewise, as long as $y_{S,T}^{(c)}$ are not too large at the weak scale, couplings do not run non-perturbative.

In the left panels of \Fig{fig:S_mh_scalar}, \Fig{fig:T_mh_scalar}, and \Fig{fig:D_mh_scalar} we present metastability bands in the $(\Lambda,m_H)$ plane for the singlet, triplet, and doublet scalar theories, respectively.  Here we have fixed $m_S = 100$ GeV, $m_T = 200$ GeV, and $m_D = 500$ GeV for the one-loop threshold corrections.  Meanwhile, we have assumed vanishing self-quartic couplings at the weak scale, $\lambda_S = \lambda_T = \lambda_D = \lambda_{D}' =0$, whose effects on Higgs stability are subleading, while the cross-quartic couplings have been allowed to vary.  For the doublet scalar theory we have also fixed $\kappa_D'=0$ for simplicity. Regions corresponding to an unstable vacuum, \emph{i.e.}~$\lambda_H < \hat \lambda_H$, are not shown because these occur in a light Higgs range which is already experimentally excluded.

From these plots we see that additional cross-quartic couplings to the Higgs tend to stabilize the vacuum.  This occurs because scalar loops drive the Higgs quartic positive in the ultraviolet.  Thus, while \Fig{fig:SM} indicates that $m_H \simeq 125$ GeV implies a stability cutoff of $\Lambda\simeq 10^{10}$ GeV, this can be lifted up to the GUT or Planck scale, $\Lambda \simeq 10^{16} - 10^{19}$ GeV, if the cross-quartic couplings are evenly modestly sized.
Finally, note that in these plots all of the theory parameters remain fully perturbative, primarily because the weak scale values of the cross-quartic couplings were chosen to be small.

On the other hand, the right panels of \Fig{fig:S_mh_scalar}, \Fig{fig:T_mh_scalar}, and \Fig{fig:D_mh_scalar} depict the scale $\Lambda$ at which any one of the theory parameters become non-perturbative.  We use the simple criterion for perturbativity outlined in \Sec{sec:methodology}.  We see that perturbativity by itself places a significant constraint on the scale where new physics must enter, especially for larger couplings.

\begin{figure}[htbp]
\hspace*{-1cm}
\includegraphics[scale=1.]{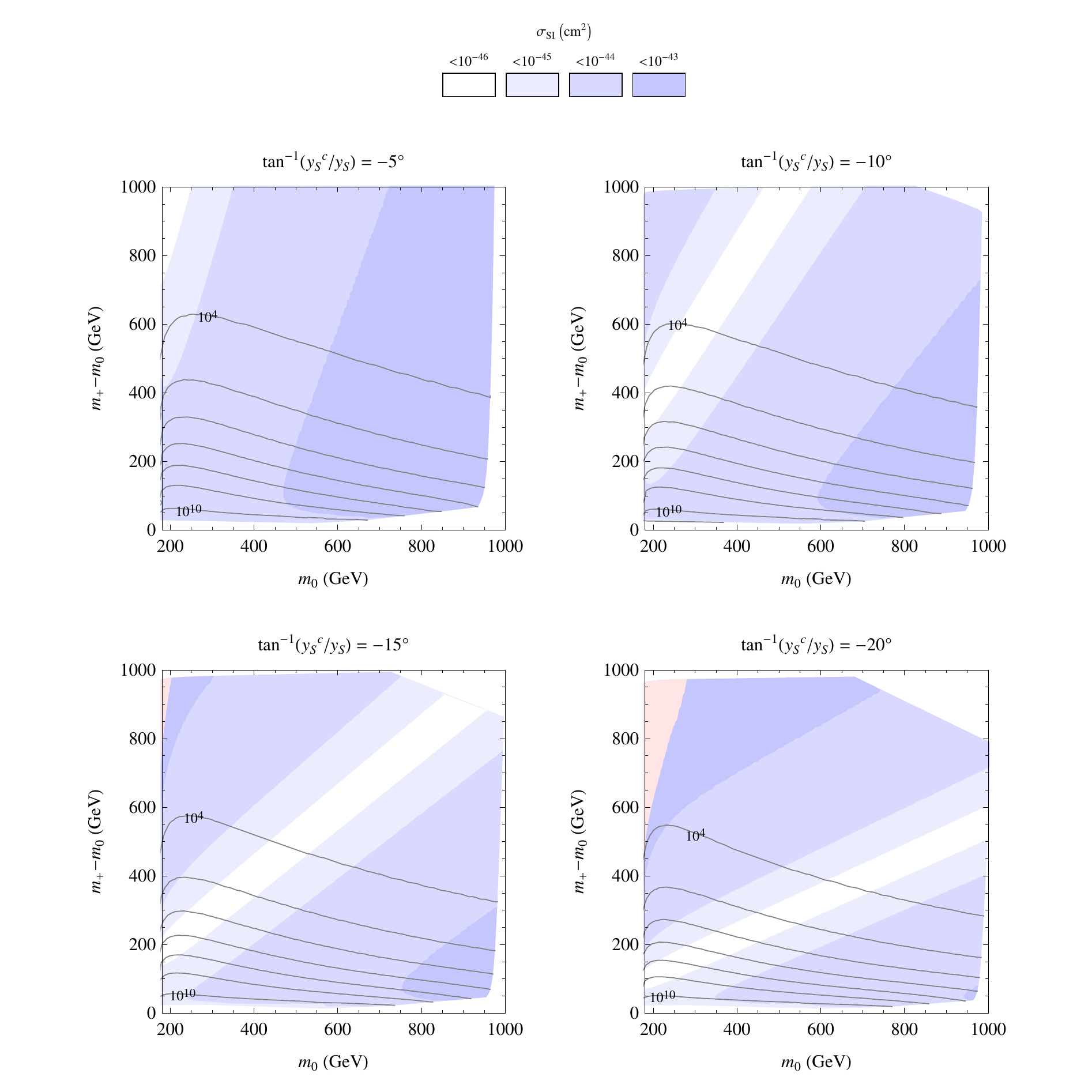}
\caption{Contour plots for SM + singlet/doublet fermion with singlet-like DM, shown in the $(m_0, m_+-m_0)$ plane at fixed values of $y_S^c/y_S$.  All points are consistent with $\Omega h^2 = 0.11$, while the red/blue regions are excluded/allowed by XENON100, with $\sigma_{SI}$ denoted.  The solid gray contours denote the scale $\Lambda$ (GeV) at which the vacuum becomes metastable. \label{fig:angle_plot} }\end{figure}

\begin{figure}[htbp]
\hspace*{-1cm}
\includegraphics[scale=1]{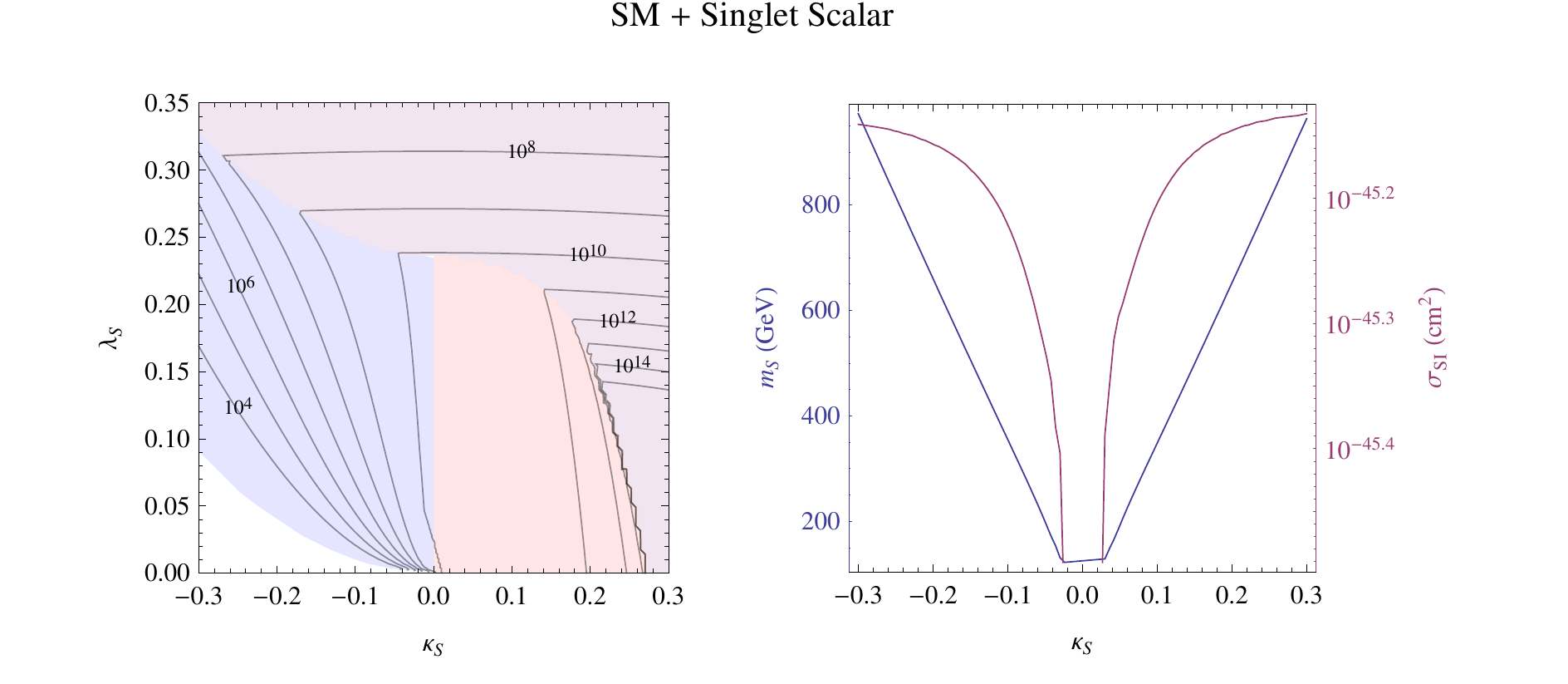}
\caption{Plots for SM + singlet scalar DM, shown as a function of the quartic couplings, $\kappa_S$ and $\lambda_S$. The left panel depicts contours $\Lambda$ (GeV) required by stability and perturbativity.  The red/blue/purple regions correspond to regimes in which $\Lambda$ is determined by metastability due to a negative Higgs quartic/metastability due to a large and negative cross-quartic/non-perturbative couplings.  The right panel depicts the value of $m_S$ required for a thermal relic abundance along with the predicted $\sigma_{SI}$.\label{fig:S_scalar}}
\end{figure}

\begin{figure}[htbp]
\hspace*{-1cm}
\includegraphics[scale=1]{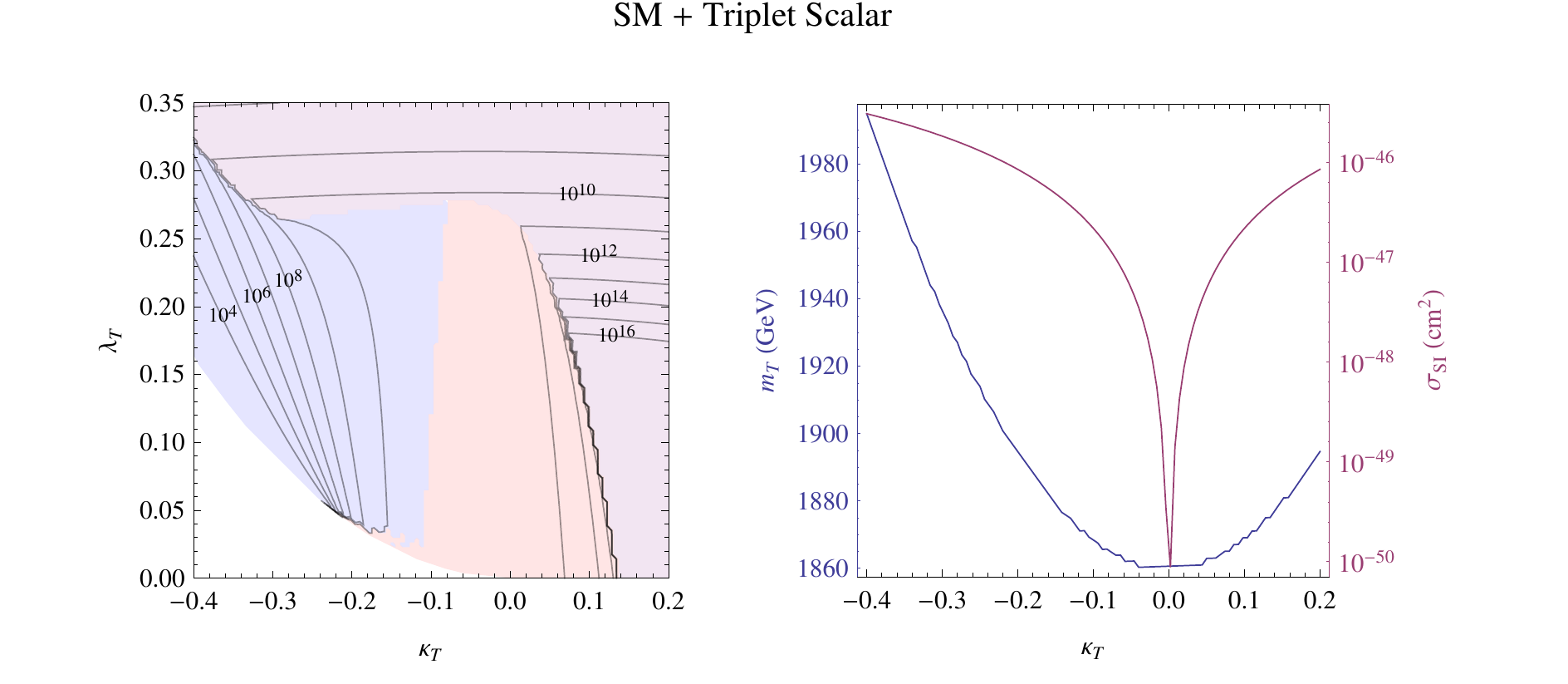}
\caption{Same as \Fig{fig:S_scalar} but for SM + triplet scalar DM.\label{fig:T_scalar}}
\end{figure}

\begin{figure}[htbp]
\hspace*{-1cm}
\includegraphics[scale=1]{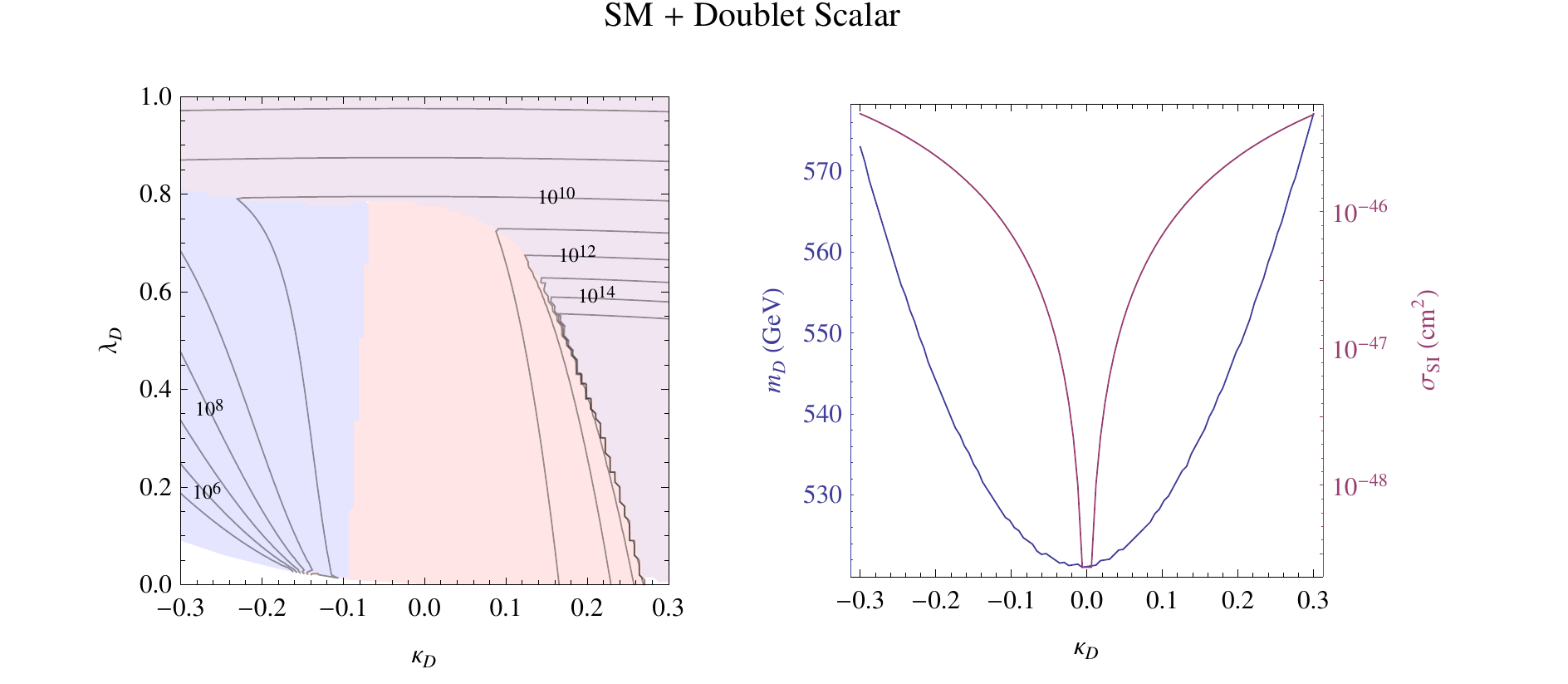}
\caption{Same as \Fig{fig:S_scalar} but for SM + doublet scalar DM.\label{fig:D_scalar}}
\end{figure} 

\subsection{Stability, Perturbativity and Dark Matter}

Thus far we have neglected the fact that the states under consideration should comprise a viable thermal relic DM candidate consistent with current direct detection constraints.  Imposing these additional criteria reduces the theory parameter space substantially, allowing for more conclusive statements.  

To begin, let us consider the case of the singlet/doublet fermion theory represented in \Fig{fig:SD_fermion_S} in the $(y_S^c, y_S)$ plane. Different values for $m_S$ and $m_D$ are shown in different plots.  Because we have chosen $m_S < m_D$, the DM is singlet-like.  Each point in this plot corresponds to a different choice of model parameters, for which we can evaluate stability and check consistency with respect to a thermal DM relic abundance and current constraints on $\sigma_{SI}$, the spin independent DM scattering cross-section, from XENON100.  The red/blue regions are excluded/allowed by XENON100, while the purple bands correspond to the WMAP favored region, $\Omega h^2 = 0.11 \pm 0.01$.  As denoted by the key, the gradations of blue denote different values for the spin independent direct detection cross-section, $\sigma_{SI}$.  The labeled contours in the upper left/lower right triangles denote the lower/upper values of $\Lambda$ which border the metastability band.  Said another way, the contours in the upper left triangle correspond to $\lambda_H =0$ while those in the lower right correspond to $\lambda_H = \hat \lambda_H$.

The shape of the XENON100 excluded regions in red can be understood as follows.  As the Yukawas increase, the coupling between the Higgs boson and DM tends to increase as well, modulo the effects of mixing.  Consequently the regions excluded from constraints on $\sigma_{SI}$ typically occur at larger values of $y_S^{(c)}$.  An important caveat, however, is that the Higgs coupling to DM is strongly suppressed  for certain values of $y_S/y_S^c$, since they control the mixing angles. This effect is responsible of the allowed regions in blue extending to the boundaries of the plot.  This feature was discussed in the case of general mixed singlet/doublet DM in~\cite{PierceCohen}.

Likewise, the shape of the WMAP favored region is clear because at $y_S = \pm y_S^c$, the coupling between the $Z$ boson and DM vanishes.  Hence, DM annihilation channels involving the $Z$ boson are closed.  In order to boost the annihilation cross-section, the Yukawa couplings must then increase, enhancing annihilations to Higgs bosons.    

Finally, let us address the contours characterizing vacuum stability.  As expected, as $y_S$ and $y_S^c$ increase, the Higgs quartic is driven negative in the ultraviolet and so the vacuum grows more and more unstable.  Because metastability is a less stringent criterion than absolute stability, the curves in the lower right triangle are in general larger than those in the upper left triangle.  From the plots it is clear that fermionic DM produced from thermal freeze-out and consistent with XENON100 tends to destabilize the electroweak symmetry breaking vacuum.  In particular, the vacuum becomes metastable in some cases even below $\Lambda \lesssim 10^{4}$ GeV.  The fact that the cutoff $\Lambda$ of this theory may be so low indicates that in these models new physics may be lurking in an energy range relevant for indirect searches in flavor or CP violating interactions.  Note that if the stable fermion is only a subdominant component of the DM in the Universe, then the purple band is a lower bound on the couplings $y_S$ and $y_S^c$, in which case the vacuum will be destabilized even more.

\Fig{fig:SD_fermion_D} is the same as \Fig{fig:SD_fermion_S} except $m_D < m_S$, so DM is doublet-like.  Likewise, \Fig{fig:TD_fermion_T} and \Fig{fig:TD_fermion_D} are the same as \Fig{fig:SD_fermion_S} only for the doublet/triplet fermion theory.  In  \Fig{fig:TD_fermion_T}, we have chosen $m_T < m_D$ to yield triplet-like DM, while  \Fig{fig:TD_fermion_D} corresponds to $m_D < m_T$ and thus doublet-like DM.

Next, consider \Fig{fig:angle_plot}, in which we present contour plots in the $(m_0, m_+-m_0)$ plane for the singlet/doublet fermion theory.  Each panel corresponds to a different value of the mixing angle, $y_S^c/y_S$, and we have again fixed $m_S < m_D$ so that DM is singlet-like.  Every point on this plot is consistent with the WMAP favored region, $\Omega h^2 = 0.11$, thus fixing one combination of the four theory parameters $\{m_S, m_D, y_S, y_S^c\}$.  We have then plotted contours of the scale $\Lambda$ at which the vacuum becomes metastable, as well as contours of $\sigma_{SI}$.  As before, the red/blue regions indicate parameter space which is excluded/allowed by XENON100.

Some remarks are in order.  As we saw in \Fig{fig:SD_fermion_S}, at certain values of $y_S^c/y_S$, there is a cancellation in the coupling of the Higgs boson to DM due to mixing, and $\sigma_{SI}$ drops to zero. In practice there will still be higher order contributions to $\sigma_{SI}$, but they will be highly suppressed and therefore we neglected to present them here. On the other hand, in these regions of parameter space the spin-dependent cross-section, $\sigma_{SD}$ will be non-zero, since it is controlled by $Z$ boson exchange.  In particular, $\sigma_{SD}$ lies in a range of $10^{-39}$ - $10^{-41}$ cm$^{2}$, accessible to experiments in the upcoming future. The same feature appears in \Fig{fig:angle_plot}, where the white regions corresponds to the point of maximal cancellation in the direct detection scattering cross-section.  From this plot, it is clear that the next generation of direct detection experiments will limit a large portion of the allowed region.  

Regarding the stability bounds, one sees that as $m_+ - m_0$ increases in value, then the singlet and doublet fermion components in general become less mixed.  In order that the singlet-like DM not be overly abundant, $y_S$ and $y_S^c$ must increase, which in turn destabilizes the vacuum.  As can be seen from \Fig{fig:angle_plot}, the vacuum can become metastable at scales as low as $10^4$ GeV in the case of large $m_+ - m_0$.

Lastly, let us now consider the case of scalar DM.  We begin with \Fig{fig:S_scalar}, in which we present results for the SM + singlet scalar DM.   In the left panel of \Fig{fig:S_scalar}, each colored region depicts a distinct phase of the theory.  For any given point in parameter space, a cutoff $\Lambda$ may be required due to some pathology---either because a stability condition fails for the Higgs quartic ($\lambda_H < 0$), a stability condition fails for the cross-quartic ($\kappa_S <  -2\sqrt{ \lambda_H \lambda_S}$), or a coupling becomes non-perturbative.  These regimes are indicated in the left panel by the blue, red, and purple regions, respectively.  Meanwhile, the labeled contours indicate the lowest cutoff dictated by any of these criteria failing.    Naturally, the stability bound on the cross-quartic dominates for negative values of $\kappa_S$, while the perturbativity bound dominates for large values of $\lambda_S$.    In the right panel of  \Fig{fig:S_scalar} we have plotted the critical value for $m_S$ required to yield the WMAP favored relic abundance, $\Omega h^2 = 0.11$, as function of the cross-quartic, $\kappa_S$. These values for $m_S$ were used to compute the one-loop threshold corrections in \Fig{fig:S_scalar}.  \Fig{fig:T_scalar} and \Fig{fig:D_scalar} are the same, but for SM + triplet and doublet DM.

\section{Conclusions}

In this work we have studied the effect of weak scale DM on vacuum stability and perturbativity.   Our primary result is that theories with thermal relic DM coupling via the Higgs  typically require new dynamics {\it in addition to} DM at a scale $\Lambda$ well below the Planck scale.  For heavy fermionic DM, $\Lambda$ can easily be as low as 10 - 1000 TeV as a consequence of vacuum stability.  On the other hand, in theories of scalar DM, $\Lambda$ takes a broad range, depending  largely on the criteria of perturbativity as well as stability.  Because $\Lambda$ is correlated with the mass and scattering cross-section of DM, direct detection experiments may serve as an indirect probe of the scale of new physics.
In short, while new physics at the scale $\Lambda$ is likely inaccessible to the LHC, indirect effects may be probed in the cosmology frontier via direct detection experiments, or in the intensity frontier via high luminosity probes.
Lastly, note that the new dynamics introduced is often required to couple to the Higgs---\emph{e.g.}~if its purpose is to prevent vacuum decay---but  it may or may not interact directly with SM fermions.  In such cases, experiments probing flavor conserving CP violation may be more likely venues for indirect signals than those considering flavor violating phenomena.

\vspace*{0.5cm}

\noindent {\bf Acknowledgments}  C.C. and M.P.~are supported in part by the Director, Office of Science, Office of High Energy and Nuclear Physics, of the US Department of Energy under Contract DE-AC02-05CH11231. C.C. is also supported in part by the National Science Foundation under grant PHY-0855653.  K.Z.~is supported by NSF CAREER award PHY1049896 and NASA Astrophysics Theory grant NNX11AI17G.   K.Z. thanks the Berkeley theory group for hospitality where this work was initiated.

\appendix

\section{Two-Loop Renormalization Group Equations}
\label{app:RGEs}

We define the following convenient notation in which the RGEs for a given coupling strength $g$ are given by
\bea
\frac{d g}{dt} &=& \frac{b^{(1)}_g}{(4\pi)^2} + \frac{b^{(2)}_g}{(4\pi)^4}, 
\eea
where $t = \log \left(Q/Q_{0}\right)$ and $b^{(1)}_g$ and $b^{(2)}_g$ are the one- and two-loop beta functions for $g$, respectively.  For thoroughness we also recall the well-known RGEs for the SM, defined in the notation of \Eq{eq:SMnotation}. Afterwards, we present results for new physics models, which induce contributions of the form
\bea
b^{(1)}_g & \rightarrow&  b^{(1)}_g + \Delta b^{(1)}_g\\
b^{(2)}_g  &\rightarrow&  b^{(1)}_g + \Delta b^{(1)}_g,
 \eea
 for each coupling $g$.  When the expression $\Delta b^{(1),(2)}_g$ is not shown, then the corresponding one- or two-loop quantity receives no corrections from new physics. The RGEs for the fermionic models were derived independently and presented in~\cite{Giudice:2011cg}.

\subsection{SM}

\begin{changemargin}{1.22cm}{0cm}

\hspace*{-1.9cm}  $ b^{(1)}_{g_1} =\frac{41 g_1^3}{10}$ \vspace*{0.25cm}

\hspace*{-1.9cm}  $ b^{(2)}_{g_1} =-\frac{1}{2} g_1^3 y_b^2-\frac{17}{10} g_1^3 y_t^2-\frac{3}{2} g_1^3 y_{\tau }^2+\frac{199 g_1^5}{50}+\frac{27}{10} g_2^2 g_1^3+\frac{44}{5} g_3^2 g_1^3$ \vspace*{0.25cm}

\hspace*{-1.9cm}  $ b^{(1)}_{g_2} =-\frac{19 g_2^3}{6}$ \vspace*{0.25cm}

\hspace*{-1.9cm}  $ b^{(2)}_{g_2} =-\frac{3}{2} g_2^3 y_b^2-\frac{3}{2} g_2^3 y_t^2-\frac{1}{2} g_2^3 y_{\tau }^2+\frac{35 g_2^5}{6}+\frac{9}{10} g_1^2 g_2^3+12 g_3^2 g_2^3$ \vspace*{0.25cm}

\hspace*{-1.9cm}  $ b^{(1)}_{g_3} =-7 g_3^3$ \vspace*{0.25cm}

\hspace*{-1.9cm}  $ b^{(2)}_{g_3} =-2 g_3^3 y_b^2-2 g_3^3 y_t^2-26 g_3^5+\frac{11}{10} g_1^2 g_3^3+\frac{9}{2} g_2^2 g_3^3$ \vspace*{0.25cm}

\hspace*{-1.9cm}  $ b^{(1)}_{y_t} =\frac{3}{2} y_b^2 y_t-\frac{17}{20} g_1^2 y_t-\frac{9}{4} g_2^2 y_t-8 g_3^2 y_t+y_t y_{\tau }^2+\frac{9 y_t^3}{2}$ \vspace*{0.25cm}

\hspace*{-1.9cm}  $ b^{(2)}_{y_t} =\frac{7}{80} g_1^2 y_b^2 y_t+\frac{99}{16} g_2^2 y_b^2 y_t+4 g_3^2 y_b^2 y_t+\frac{5}{4} y_b^2 y_t y_{\tau }^2-\frac{11}{4} y_b^2 y_t^3-\frac{1}{4} y_b^4 y_t+\frac{15}{8} g_1^2 y_t y_{\tau }^2+\frac{15}{8} g_2^2 y_t y_{\tau }^2+\frac{393}{80} g_1^2 y_t^3+\frac{225}{16} g_2^2 y_t^3+36 g_3^2 y_t^3+\frac{1187}{600} g_1^4 y_t-\frac{23}{4} g_2^4 y_t-108 g_3^4 y_t-\frac{9}{20} g_1^2 g_2^2 y_t+\frac{19}{15} g_1^2 g_3^2 y_t+9 g_2^2 g_3^2 y_t-6 \lambda _H y_t^3+\frac{3}{2} \lambda _H^2 y_t-\frac{9}{4} y_t^3 y_{\tau }^2-\frac{9}{4} y_t y_{\tau }^4-12 y_t^5$ \vspace*{0.25cm}

\hspace*{-1.9cm}  $ b^{(1)}_{y_b} =-\frac{1}{4} g_1^2 y_b-\frac{9}{4} g_2^2 y_b-8 g_3^2 y_b+\frac{3}{2} y_b y_t^2+y_b y_{\tau }^2+\frac{9 y_b^3}{2}$ \vspace*{0.25cm}

\hspace*{-1.9cm}  $ b^{(2)}_{y_b} =\frac{91}{80} g_1^2 y_b y_t^2+\frac{99}{16} g_2^2 y_b y_t^2+4 g_3^2 y_b y_t^2+\frac{15}{8} g_1^2 y_b y_{\tau }^2+\frac{15}{8} g_2^2 y_b y_{\tau }^2+\frac{237}{80} g_1^2 y_b^3+\frac{225}{16} g_2^2 y_b^3+36 g_3^2 y_b^3-\frac{127}{600} g_1^4 y_b-\frac{23}{4} g_2^4 y_b-108 g_3^4 y_b-\frac{27}{20} g_1^2 g_2^2 y_b+\frac{31}{15} g_1^2 g_3^2 y_b+9 g_2^2 g_3^2 y_b-6 y_b^3 \lambda _H+\frac{3}{2} y_b \lambda _H^2+\frac{5}{4} y_b y_t^2 y_{\tau }^2-\frac{11}{4} y_b^3 y_t^2-\frac{1}{4} y_b y_t^4-\frac{9}{4} y_b^3 y_{\tau }^2-\frac{9}{4} y_b y_{\tau }^4-12 y_b^5$ \vspace*{0.25cm}

\hspace*{-1.9cm}  $ b^{(1)}_{y_{\tau }} =3 y_b^2 y_{\tau }-\frac{9}{4} g_1^2 y_{\tau }-\frac{9}{4} g_2^2 y_{\tau }+3 y_t^2 y_{\tau }+\frac{5 y_{\tau }^3}{2}$ \vspace*{0.25cm}

\hspace*{-1.9cm}  $ b^{(2)}_{y_{\tau }} =\frac{5}{8} g_1^2 y_b^2 y_{\tau }+\frac{45}{8} g_2^2 y_b^2 y_{\tau }+20 g_3^2 y_b^2 y_{\tau }+\frac{3}{2} y_b^2 y_t^2 y_{\tau }-\frac{27}{4} y_b^2 y_{\tau }^3-\frac{27}{4} y_b^4 y_{\tau }+\frac{17}{8} g_1^2 y_t^2 y_{\tau }+\frac{45}{8} g_2^2 y_t^2 y_{\tau }+20 g_3^2 y_t^2 y_{\tau }+\frac{537}{80} g_1^2 y_{\tau }^3+\frac{165}{16} g_2^2 y_{\tau }^3+\frac{1371}{200} g_1^4 y_{\tau }-\frac{23}{4} g_2^4 y_{\tau }+\frac{27}{20} g_1^2 g_2^2 y_{\tau }-6 \lambda _H y_{\tau }^3+\frac{3}{2} \lambda _H^2 y_{\tau }-\frac{27}{4} y_t^2 y_{\tau }^3-\frac{27}{4} y_t^4 y_{\tau }-3 y_{\tau }^5$ \vspace*{0.25cm}

\hspace*{-1.9cm}  $ b^{(1)}_{\lambda _H} =12 y_b^2 \lambda _H-12 y_b^4-\frac{9}{5} g_1^2 \lambda _H-9 g_2^2 \lambda _H+\frac{27 g_1^4}{100}+\frac{9}{10} g_2^2 g_1^2+\frac{9 g_2^4}{4}+12 \lambda _H y_t^2+4 \lambda _H y_{\tau }^2+12 \lambda _H^2-12 y_t^4-4 y_{\tau }^4$ \vspace*{0.25cm}

\hspace*{-1.9cm}  $ b^{(2)}_{\lambda _H} =\frac{5}{2} g_1^2 y_b^2 \lambda _H+\frac{45}{2} g_2^2 y_b^2 \lambda _H+80 g_3^2 y_b^2 \lambda _H+\frac{9}{10} g_1^4 y_b^2+\frac{8}{5} g_1^2 y_b^4+\frac{27}{5} g_2^2 g_1^2 y_b^2-64 g_3^2 y_b^4-\frac{9}{2} g_2^4 y_b^2-42 y_b^2 \lambda _H y_t^2-72 y_b^2 \lambda _H^2-3 y_b^4 \lambda _H-12 y_b^2 y_t^4-12 y_b^4 y_t^2+60 y_b^6+\frac{17}{2} g_1^2 \lambda _H y_t^2+\frac{45}{2} g_2^2 \lambda _H y_t^2+80 g_3^2 \lambda _H y_t^2+\frac{15}{2} g_1^2 \lambda _H y_{\tau }^2+\frac{15}{2} g_2^2 \lambda _H y_{\tau }^2+\frac{1887}{200} g_1^4 \lambda _H+\frac{54}{5} g_1^2 \lambda _H^2+\frac{117}{20} g_2^2 g_1^2 \lambda _H+54 g_2^2 \lambda _H^2-\frac{73}{8} g_2^4 \lambda _H-\frac{171}{50} g_1^4 y_t^2-\frac{16}{5} g_1^2 y_t^4+\frac{63}{5} g_2^2 g_1^2 y_t^2-64 g_3^2 y_t^4-\frac{9}{2} g_2^4 y_t^2-\frac{9}{2} g_1^4 y_{\tau }^2-\frac{24}{5} g_1^2 y_{\tau }^4+\frac{33}{5} g_2^2 g_1^2 y_{\tau }^2-\frac{3}{2} g_2^4 y_{\tau }^2-\frac{3411 g_1^6}{1000}-\frac{1677}{200} g_2^2 g_1^4-\frac{289}{40} g_2^4 g_1^2+\frac{305 g_2^6}{8}-72 \lambda _H^2 y_t^2-3 \lambda _H y_t^4-24 \lambda _H^2 y_{\tau }^2-\lambda _H y_{\tau }^4-78 \lambda _H^3+60 y_t^6+20 y_{\tau }^6$ \vspace*{0.25cm}

\end{changemargin}

\subsection{SM + Real Singlet Scalar}

\begin{changemargin}{1.22cm}{0cm}

\hspace*{-1.9cm}  $ \Delta b^{(2)}_{y_t} =\frac{1}{4} \kappa _S^2 y_t$ \vspace*{0.25cm}

\hspace*{-1.9cm}  $ \Delta b^{(2)}_{y_b} =\frac{1}{4} y_b \kappa _S^2$ \vspace*{0.25cm}

\hspace*{-1.9cm}  $ \Delta b^{(2)}_{y_{\tau }} =\frac{1}{4} \kappa _S^2 y_{\tau }$ \vspace*{0.25cm}

\hspace*{-1.9cm}  $ \Delta b^{(1)}_{\lambda _H} =\kappa _S^2$ \vspace*{0.25cm}

\hspace*{-1.9cm}  $ \Delta b^{(2)}_{\lambda _H} =-5 \lambda _H \kappa _S^2-4 \kappa _S^3$ \vspace*{0.25cm}

\hspace*{-1.9cm}  $ b^{(1)}_{\lambda _S} =\kappa _S^2+36 \lambda _S^2$ \vspace*{0.25cm}

\hspace*{-1.9cm}  $ b^{(2)}_{\lambda _S} =-6 y_b^2 \kappa _S^2+\frac{6}{5} g_1^2 \kappa _S^2+6 g_2^2 \kappa _S^2-6 \kappa _S^2 y_t^2-2 \kappa _S^2 y_{\tau }^2-20 \kappa _S^2 \lambda _S-4 \kappa _S^3-816 \lambda _S^3$ \vspace*{0.25cm}

\hspace*{-1.9cm}  $ b^{(1)}_{\kappa _S} =6 y_b^2 \kappa _S-\frac{9}{10} g_1^2 \kappa _S-\frac{9}{2} g_2^2 \kappa _S+6 \lambda _H \kappa _S+6 \kappa _S y_t^2+2 \kappa _S y_{\tau }^2+12 \kappa _S \lambda _S+4 \kappa _S^2$ \vspace*{0.25cm}

\hspace*{-1.9cm}  $ b^{(2)}_{\kappa _S} =\frac{5}{4} g_1^2 y_b^2 \kappa _S+\frac{45}{4} g_2^2 y_b^2 \kappa _S+40 g_3^2 y_b^2 \kappa _S-36 y_b^2 \lambda _H \kappa _S-21 y_b^2 \kappa _S y_t^2-12 y_b^2 \kappa _S^2-\frac{27}{2} y_b^4 \kappa _S+\frac{36}{5} g_1^2 \lambda _H \kappa _S+36 g_2^2 \lambda _H \kappa _S+\frac{17}{4} g_1^2 \kappa _S y_t^2+\frac{45}{4} g_2^2 \kappa _S y_t^2+40 g_3^2 \kappa _S y_t^2+\frac{15}{4} g_1^2 \kappa _S y_{\tau }^2+\frac{15}{4} g_2^2 \kappa _S y_{\tau }^2+\frac{1671}{400} g_1^4 \kappa _S+\frac{3}{5} g_1^2 \kappa _S^2+\frac{9}{8} g_2^2 g_1^2 \kappa _S+3 g_2^2 \kappa _S^2-\frac{145}{16} g_2^4 \kappa _S-36 \lambda _H \kappa _S y_t^2-12 \lambda _H \kappa _S y_{\tau }^2-15 \lambda _H^2 \kappa _S-36 \lambda _H \kappa _S^2-12 \kappa _S^2 y_t^2-\frac{27}{2} \kappa _S y_t^4-4 \kappa _S^2 y_{\tau }^2-\frac{9}{2} \kappa _S y_{\tau }^4-120 \kappa _S \lambda _S^2-72 \kappa _S^2 \lambda _S-\frac{21 \kappa _S^3}{2}$ \vspace*{0.25cm}

\end{changemargin}
\subsection{SM + Real Triplet Scalar.}

\begin{changemargin}{1.22cm}{0cm}

\hspace*{-1.9cm}  $ \Delta b^{(1)}_{g_2} =\frac{g_2^3}{3}$ \vspace*{0.25cm}

\hspace*{-1.9cm}  $ \Delta b^{(2)}_{g_2} =\frac{28 g_2^5}{3}$ \vspace*{0.25cm}

\hspace*{-1.9cm}  $ \Delta b^{(2)}_{y_t} =g_2^4 y_t+\frac{3}{4} y_t \kappa _T^2$ \vspace*{0.25cm}

\hspace*{-1.9cm}  $ \Delta b^{(2)}_{y_b} =g_2^4 y_b+\frac{3}{4} y_b \kappa _T^2$ \vspace*{0.25cm}

\hspace*{-1.9cm}  $ \Delta b^{(2)}_{y_{\tau }} =g_2^4 y_{\tau }+\frac{3}{4} \kappa _T^2 y_{\tau }$ \vspace*{0.25cm}

\hspace*{-1.9cm}  $ \Delta b^{(1)}_{\lambda _H} =3 \kappa _T^2$ \vspace*{0.25cm}

\hspace*{-1.9cm}  $ \Delta b^{(2)}_{\lambda _H} =\frac{11}{2} g_2^4 \lambda _H+30 g_2^4 \kappa _T+48 g_2^2 \kappa _T^2-\frac{7 g_2^6}{2}-\frac{7}{10} g_1^2 g_2^4-15 \lambda _H \kappa _T^2-12 \kappa _T^3$ \vspace*{0.25cm}

\hspace*{-1.9cm}  $ b^{(1)}_{\lambda _T} =-24 g_2^2 \lambda _T+6 g_2^4+\kappa _T^2+44 \lambda _T^2$ \vspace*{0.25cm}

\hspace*{-1.9cm}  $ b^{(2)}_{\lambda _T} =-6 y_b^2 \kappa _T^2+10 g_2^4 \kappa _T+6 g_2^2 \kappa _T^2+\frac{6}{5} g_1^2 \kappa _T^2+\frac{382}{3} g_2^4 \lambda _T+512 g_2^2 \lambda _T^2-\frac{110 g_2^6}{3}-6 y_t^2 \kappa _T^2-2 \kappa _T^2 y_{\tau }^2-20 \kappa _T^2 \lambda _T-4 \kappa _T^3-1104 \lambda _T^3$ \vspace*{0.25cm}

\hspace*{-1.9cm}  $ b^{(1)}_{\kappa _T} =6 y_b^2 \kappa _T-\frac{33}{2} g_2^2 \kappa _T-\frac{9}{10} g_1^2 \kappa _T+6 g_2^4+6 \lambda _H \kappa _T+6 y_t^2 \kappa _T+2 \kappa _T y_{\tau }^2+20 \kappa _T \lambda _T+4 \kappa _T^2$ \vspace*{0.25cm}

\hspace*{-1.9cm}  $ b^{(2)}_{\kappa _T} =\frac{45}{4} g_2^2 y_b^2 \kappa _T+\frac{5}{4} g_1^2 y_b^2 \kappa _T+40 g_3^2 y_b^2 \kappa _T-6 g_2^4 y_b^2-36 y_b^2 \lambda _H \kappa _T-21 y_b^2 y_t^2 \kappa _T-12 y_b^2 \kappa _T^2-\frac{27}{2} y_b^4 \kappa _T+36 g_2^2 \lambda _H \kappa _T+\frac{36}{5} g_1^2 \lambda _H \kappa _T+30 g_2^4 \lambda _H+\frac{45}{4} g_2^2 y_t^2 \kappa _T+\frac{17}{4} g_1^2 y_t^2 \kappa _T+40 g_3^2 y_t^2 \kappa _T-6 g_2^4 y_t^2+\frac{15}{4} g_2^2 \kappa _T y_{\tau }^2+\frac{15}{4} g_1^2 \kappa _T y_{\tau }^2+320 g_2^2 \kappa _T \lambda _T+\frac{641}{48} g_2^4 \kappa _T+11 g_2^2 \kappa _T^2+\frac{9}{8} g_1^2 g_2^2 \kappa _T+\frac{3}{5} g_1^2 \kappa _T^2+\frac{1671}{400} g_1^4 \kappa _T+100 g_2^4 \lambda _T-2 g_2^4 y_{\tau }^2+\frac{329 g_2^6}{6}-\frac{9}{2} g_1^2 g_2^4-36 \lambda _H y_t^2 \kappa _T-12 \lambda _H \kappa _T y_{\tau }^2-15 \lambda _H^2 \kappa _T-36 \lambda _H \kappa _T^2-12 y_t^2 \kappa _T^2-\frac{27}{2} y_t^4 \kappa _T-4 \kappa _T^2 y_{\tau }^2-\frac{9}{2} \kappa _T y_{\tau }^4-200 \kappa _T \lambda _T^2-120 \kappa _T^2 \lambda _T-\frac{23 \kappa _T^3}{2}$ \vspace*{0.25cm}

\end{changemargin}

\subsection{SM + Complex Doublet Scalar}

\begin{changemargin}{1.22cm}{0cm}

\hspace*{-1.9cm}  $ \Delta b^{(1)}_{g_1} =\frac{g_1^3}{10}$ \vspace*{0.25cm}

\hspace*{-1.9cm}  $ \Delta b^{(2)}_{g_1} =\frac{9 g_1^5}{50}+\frac{9}{10} g_2^2 g_1^3$ \vspace*{0.25cm}

\hspace*{-1.9cm}  $ \Delta b^{(1)}_{g_2} =\frac{g_2^3}{6}$ \vspace*{0.25cm}

\hspace*{-1.9cm}  $ \Delta b^{(2)}_{g_2} =\frac{13 g_2^5}{6}+\frac{3}{10} g_1^2 g_2^3$ \vspace*{0.25cm}

\hspace*{-1.9cm}  $ \Delta b^{(2)}_{y_t} =\frac{1}{4} {\kappa_D'}{} \kappa _D y_t+\frac{1}{4} \kappa _D^2 y_t+\frac{2}{15} g_1^4 y_t+\frac{1}{2} g_2^4 y_t+\frac{{\kappa_D'}{}^2 y_t}{4}$ \vspace*{0.25cm}

\hspace*{-1.9cm}  $ \Delta b^{(2)}_{y_b} =\frac{1}{4} {\kappa_D'}{} y_b \kappa _D+\frac{1}{4} y_b \kappa _D^2+\frac{7}{300} g_1^4 y_b+\frac{1}{2} g_2^4 y_b+\frac{{\kappa_D'}{}^2 y_b}{4}$ \vspace*{0.25cm}

\hspace*{-1.9cm}  $ \Delta b^{(2)}_{y_{\tau }} =\frac{1}{4} {\kappa_D'}{} \kappa _D y_{\tau }+\frac{1}{4} \kappa _D^2 y_{\tau }+\frac{39}{100} g_1^4 y_{\tau }+\frac{1}{2} g_2^4 y_{\tau }+\frac{{\kappa_D'}{}^2 y_{\tau }}{4}$ \vspace*{0.25cm}

\hspace*{-1.9cm}  $ \Delta b^{(1)}_{\lambda _H} ={\kappa_D'}{} \kappa _D+\kappa _D^2+\frac{{\kappa_D'}{}^2}{2}$ \vspace*{0.25cm}

\hspace*{-1.9cm}  $ \Delta b^{(2)}_{\lambda _H} =\frac{6}{5} g_1^2 {\kappa_D'}{} \kappa _D+6 g_2^2 {\kappa_D'}{} \kappa _D+\frac{9}{10} g_1^4 \kappa _D+\frac{6}{5} g_1^2 \kappa _D^2+6 g_2^2 \kappa _D^2+\frac{15}{2} g_2^4 \kappa _D-5 {\kappa_D'}{} \kappa _D \lambda _H-5 \kappa _D^2 \lambda _H-4 {\kappa_D'}{}^2 \kappa _D-3 {\kappa_D'}{} \kappa _D^2-2 \kappa _D^3+\frac{33}{100} g_1^4 \lambda _H+\frac{11}{4} g_2^4 \lambda _H+\frac{3}{5} g_1^2 {\kappa_D'}{}^2+\frac{3}{2} g_2^2 {\kappa_D'}{}^2+\frac{9 g_1^4 {\kappa_D'}{}}{20}+\frac{3}{2} g_2^2 g_1^2 {\kappa_D'}{}+\frac{15 g_2^4 {\kappa_D'}{}}{4}-\frac{63 g_1^6}{500}-\frac{21}{100} g_2^2 g_1^4-\frac{7}{20} g_2^4 g_1^2-\frac{7 g_2^6}{4}-3 {\kappa_D'}{}^2 \lambda _H-\frac{3 {\kappa_D'}{}^3}{2}$ \vspace*{0.25cm}

\hspace*{-1.9cm}  $ b^{(1)}_{\lambda _D} =-\frac{9}{5} g_1^2 \lambda _D-9 g_2^2 \lambda _D+{\kappa_D'}{} \kappa _D+\kappa _D^2+12 \lambda _D^2+\frac{27 g_1^4}{100}+\frac{9}{10} g_2^2 g_1^2+\frac{9 g_2^4}{4}+\frac{{\kappa_D'}{}^2}{2}$ \vspace*{0.25cm}

\hspace*{-1.9cm}  $ b^{(2)}_{\lambda _D} =-6 {\kappa_D'}{} y_b^2 \kappa _D-6 y_b^2 \kappa _D^2-3 {\kappa_D'}{}^2 y_b^2+\frac{6}{5} g_1^2 {\kappa_D'}{} \kappa _D+6 g_2^2 {\kappa_D'}{} \kappa _D+\frac{9}{10} g_1^4 \kappa _D+\frac{6}{5} g_1^2 \kappa _D^2+6 g_2^2 \kappa _D^2+\frac{15}{2} g_2^4 \kappa _D+\frac{1953}{200} g_1^4 \lambda _D+\frac{54}{5} g_1^2 \lambda _D^2+\frac{117}{20} g_2^2 g_1^2 \lambda _D+54 g_2^2 \lambda _D^2-\frac{51}{8} g_2^4 \lambda _D-4 {\kappa_D'}{}^2 \kappa _D-3 {\kappa_D'}{}^2 \lambda _D-6 {\kappa_D'}{} \kappa _D y_t^2-2 {\kappa_D'}{} \kappa _D y_{\tau }^2-5 {\kappa_D'}{} \kappa _D \lambda _D-3 {\kappa_D'}{} \kappa _D^2-6 \kappa _D^2 y_t^2-2 \kappa _D^2 y_{\tau }^2-5 \kappa _D^2 \lambda _D-2 \kappa _D^3-78 \lambda _D^3+\frac{3}{5} g_1^2 {\kappa_D'}{}^2+\frac{3}{2} g_2^2 {\kappa_D'}{}^2+\frac{9 g_1^4 {\kappa_D'}{}}{20}+\frac{3}{2} g_2^2 g_1^2 {\kappa_D'}{}+\frac{15 g_2^4 {\kappa_D'}{}}{4}-\frac{3537 g_1^6}{1000}-\frac{1719}{200} g_2^2 g_1^4-\frac{303}{40} g_2^4 g_1^2+\frac{291 g_2^6}{8}-\frac{3 {\kappa_D'}{}^3}{2}-3 {\kappa_D'}{}^2 y_t^2-{\kappa_D'}{}^2 y_{\tau }^2$ \vspace*{0.25cm}

\hspace*{-1.9cm}  $ b^{(1)}_{\kappa _D} =6 y_b^2 \kappa _D-\frac{9}{5} g_1^2 \kappa _D-9 g_2^2 \kappa _D+6 \kappa _D \lambda _H+2 {\kappa_D'}{} \lambda _D+6 \kappa _D y_t^2+2 \kappa _D y_{\tau }^2+6 \kappa _D \lambda _D+2 \kappa _D^2+\frac{27 g_1^4}{50}-\frac{9}{5} g_2^2 g_1^2+\frac{9 g_2^4}{2}+2 {\kappa_D'}{} \lambda _H+{\kappa_D'}{}^2$ \vspace*{0.25cm}

\hspace*{-1.9cm}  $ b^{(2)}_{\kappa _D} =-\frac{3537 g_1^6}{500}+\frac{909}{100} g_2^2 g_1^4+\frac{9}{10} y_b^2 g_1^4-\frac{171}{50} y_t^2 g_1^4-\frac{9}{2} y_{\tau }^2 g_1^4+\frac{9 {\kappa_D'}{} g_1^4}{10}+\frac{1773}{200} \kappa _D g_1^4+\frac{27}{10} \lambda _D g_1^4+\frac{27}{10} \lambda _H g_1^4+\frac{33}{20} g_2^4 g_1^2-\frac{3}{5} {\kappa_D'}{}^2 g_1^2-\frac{9}{5} {\kappa_D'}{} g_2^2 g_1^2-\frac{27}{5} g_2^2 y_b^2 g_1^2-\frac{63}{5} g_2^2 y_t^2 g_1^2-\frac{33}{5} g_2^2 y_{\tau }^2 g_1^2+\frac{3}{5} \kappa _D^2 g_1^2+\frac{33}{20} g_2^2 \kappa _D g_1^2+\frac{5}{4} y_b^2 \kappa _D g_1^2+\frac{17}{4} y_t^2 \kappa _D g_1^2+\frac{15}{4} y_{\tau }^2 \kappa _D g_1^2-3 g_2^2 \lambda _D g_1^2+\frac{12}{5} {\kappa_D'}{} \lambda _D g_1^2+\frac{36}{5} \kappa _D \lambda _D g_1^2-3 g_2^2 \lambda _H g_1^2+\frac{12}{5} {\kappa_D'}{} \lambda _H g_1^2+\frac{36}{5} \kappa _D \lambda _H g_1^2+\frac{291 g_2^6}{4}+\frac{15 {\kappa_D'}{} g_2^4}{2}-3 {\kappa_D'}{}^3-3 \kappa _D^3+3 {\kappa_D'}{}^2 g_2^2-\frac{9}{2} g_2^4 y_b^2-3 {\kappa_D'}{}^2 y_b^2-\frac{9}{2} g_2^4 y_t^2-3 {\kappa_D'}{}^2 y_t^2-24 {\kappa_D'}{} y_b^2 y_t^2-\frac{3}{2} g_2^4 y_{\tau }^2-{\kappa_D'}{}^2 y_{\tau }^2+3 g_2^2 \kappa _D^2-6 y_b^2 \kappa _D^2-6 y_t^2 \kappa _D^2-2 y_{\tau }^2 \kappa _D^2-{\kappa_D'}{} \kappa _D^2-4 {\kappa_D'}{} \lambda _D^2-15 \kappa _D \lambda _D^2-4 {\kappa_D'}{} \lambda _H^2-15 \kappa _D \lambda _H^2-\frac{111}{8} g_2^4 \kappa _D-\frac{27}{2} y_b^4 \kappa _D-\frac{27}{2} y_t^4 \kappa _D-\frac{9}{2} y_{\tau }^4 \kappa _D-4 {\kappa_D'}{}^2 \kappa _D-6 {\kappa_D'}{} g_2^2 \kappa _D+\frac{45}{4} g_2^2 y_b^2 \kappa _D+40 g_3^2 y_b^2 \kappa _D+\frac{45}{4} g_2^2 y_t^2 \kappa _D+40 g_3^2 y_t^2 \kappa _D-21 y_b^2 y_t^2 \kappa _D+\frac{15}{4} g_2^2 y_{\tau }^2 \kappa _D+\frac{45}{2} g_2^4 \lambda _D-7 {\kappa_D'}{}^2 \lambda _D+18 {\kappa_D'}{} g_2^2 \lambda _D-18 \kappa _D^2 \lambda _D+36 g_2^2 \kappa _D \lambda _D-8 {\kappa_D'}{} \kappa _D \lambda _D+\frac{45}{2} g_2^4 \lambda _H-7 {\kappa_D'}{}^2 \lambda _H+18 {\kappa_D'}{} g_2^2 \lambda _H-12 {\kappa_D'}{} y_b^2 \lambda _H-12 {\kappa_D'}{} y_t^2 \lambda _H-4 {\kappa_D'}{} y_{\tau }^2 \lambda _H-18 \kappa _D^2 \lambda _H+36 g_2^2 \kappa _D \lambda _H-36 y_b^2 \kappa _D \lambda _H-36 y_t^2 \kappa _D \lambda _H-12 y_{\tau }^2 \kappa _D \lambda _H-8 {\kappa_D'}{} \kappa _D \lambda _H$ \vspace*{0.25cm}

\hspace*{-1.9cm}  $ b^{(1)}_{{\kappa_D'}{}} =6 {\kappa_D'}{} y_b^2+4 {\kappa_D'}{} \kappa _D+2 {\kappa_D'}{} \lambda _D-\frac{9 g_1^2 {\kappa_D'}{}}{5}-9 g_2^2 {\kappa_D'}{}+\frac{18}{5} g_1^2 g_2^2+2 {\kappa_D'}{} \lambda _H+2 {\kappa_D'}{}^2+6 {\kappa_D'}{} y_t^2+2 {\kappa_D'}{} y_{\tau }^2$ \vspace*{0.25cm}

\hspace*{-1.9cm}  $ b^{(2)}_{{\kappa_D'}{}} =-12 {\kappa_D'}{} y_b^2 \kappa _D+\frac{5}{4} g_1^2 {\kappa_D'}{} y_b^2+\frac{45}{4} g_2^2 {\kappa_D'}{} y_b^2+40 g_3^2 {\kappa_D'}{} y_b^2+\frac{54}{5} g_2^2 g_1^2 y_b^2-12 {\kappa_D'}{} y_b^2 \lambda _H-6 {\kappa_D'}{}^2 y_b^2+27 {\kappa_D'}{} y_b^2 y_t^2-\frac{27 {\kappa_D'}{} y_b^4}{2}+\frac{6}{5} g_1^2 {\kappa_D'}{} \kappa _D+18 g_2^2 {\kappa_D'}{} \kappa _D+\frac{12}{5} g_1^2 {\kappa_D'}{} \lambda _D+\frac{6}{5} g_2^2 g_1^2 \kappa _D+6 g_2^2 g_1^2 \lambda _D-20 {\kappa_D'}{} \kappa _D \lambda _H-7 {\kappa_D'}{}^2 \kappa _D-10 {\kappa_D'}{}^2 \lambda _D-12 {\kappa_D'}{} \kappa _D y_t^2-4 {\kappa_D'}{} \kappa _D y_{\tau }^2-20 {\kappa_D'}{} \kappa _D \lambda _D-7 {\kappa_D'}{} \kappa _D^2-7 {\kappa_D'}{} \lambda _D^2+\frac{12}{5} g_1^2 {\kappa_D'}{} \lambda _H+6 g_2^2 g_1^2 \lambda _H+\frac{12}{5} g_1^2 {\kappa_D'}{}^2+9 g_2^2 {\kappa_D'}{}^2+\frac{17}{4} g_1^2 {\kappa_D'}{} y_t^2+\frac{45}{4} g_2^2 {\kappa_D'}{} y_t^2+40 g_3^2 {\kappa_D'}{} y_t^2+\frac{15}{4} g_1^2 {\kappa_D'}{} y_{\tau }^2+\frac{15}{4} g_2^2 {\kappa_D'}{} y_{\tau }^2+\frac{1413 g_1^4 {\kappa_D'}{}}{200}+\frac{153}{20} g_2^2 g_1^2 {\kappa_D'}{}-\frac{231 g_2^4 {\kappa_D'}{}}{8}+\frac{126}{5} g_2^2 g_1^2 y_t^2+\frac{66}{5} g_2^2 g_1^2 y_{\tau }^2-\frac{657}{25} g_2^2 g_1^4-\frac{84}{5} g_2^4 g_1^2-10 {\kappa_D'}{}^2 \lambda _H-12 {\kappa_D'}{} \lambda _H y_t^2-4 {\kappa_D'}{} \lambda _H y_{\tau }^2-7 {\kappa_D'}{} \lambda _H^2-6 {\kappa_D'}{}^2 y_t^2-2 {\kappa_D'}{}^2 y_{\tau }^2-\frac{27 {\kappa_D'}{} y_t^4}{2}-\frac{9 {\kappa_D'}{} y_{\tau }^4}{2}$ \vspace*{0.25cm}

\end{changemargin}

\subsection{SM + Majorana Singlet Fermion + Dirac Doublet Fermion}

\begin{changemargin}{1.22cm}{0cm}

\hspace*{-1.9cm}  $ \Delta b^{(1)}_{g_1} =\frac{2 g_1^3}{5}$ \vspace*{0.25cm}

\hspace*{-1.9cm}  $ \Delta b^{(2)}_{g_1} =-\frac{3}{20} g_1^3 {\tilde g_d'}{}^2-\frac{3}{20} g_1^3 {\tilde g_u'}{}^2+\frac{9 g_1^5}{50}+\frac{9}{10} g_2^2 g_1^3$ \vspace*{0.25cm}

\hspace*{-1.9cm}  $ \Delta b^{(1)}_{g_2} =\frac{2 g_2^3}{3}$ \vspace*{0.25cm}

\hspace*{-1.9cm}  $ \Delta b^{(2)}_{g_2} =-\frac{1}{4} g_2^3 {\tilde g_d'}{}^2-\frac{1}{4} g_2^3 {\tilde g_u'}{}^2+\frac{49 g_2^5}{6}+\frac{3}{10} g_1^2 g_2^3$ \vspace*{0.25cm}

\hspace*{-1.9cm}  $ \Delta b^{(1)}_{y_t} =\frac{{\tilde g_d'}{}^2 y_t}{2}+\frac{{\tilde g_u'}{}^2 y_t}{2}$ \vspace*{0.25cm}

\hspace*{-1.9cm}  $ \Delta b^{(2)}_{y_t} =\frac{5}{8} {\tilde g_d'}{}^2 y_b^2 y_t+\frac{5}{8} {\tilde g_u'}{}^2 y_b^2 y_t+\frac{3}{16} g_1^2 {\tilde g_d'}{}^2 y_t+\frac{15}{16} g_2^2 {\tilde g_d'}{}^2 y_t+\frac{3}{16} g_1^2 {\tilde g_u'}{}^2 y_t+\frac{15}{16} g_2^2 {\tilde g_u'}{}^2 y_t+\frac{29}{150} g_1^4 y_t+\frac{1}{2} g_2^4 y_t-\frac{9}{16} {\tilde g_d'}{}^4 y_t-\frac{5}{4} {\tilde g_d'}{}^2 {\tilde g_u'}{}^2 y_t-\frac{9}{8} {\tilde g_d'}{}^2 y_t^3-\frac{9 {\tilde g_u'}{}^4 y_t}{16}-\frac{9}{8} {\tilde g_u'}{}^2 y_t^3$ \vspace*{0.25cm}

\hspace*{-1.9cm}  $ \Delta b^{(1)}_{y_b} =\frac{{\tilde g_d'}{}^2 y_b}{2}+\frac{{\tilde g_u'}{}^2 y_b}{2}$ \vspace*{0.25cm}

\hspace*{-1.9cm}  $ \Delta b^{(2)}_{y_b} =\frac{3}{16} g_1^2 {\tilde g_d'}{}^2 y_b+\frac{15}{16} g_2^2 {\tilde g_d'}{}^2 y_b+\frac{3}{16} g_1^2 {\tilde g_u'}{}^2 y_b+\frac{15}{16} g_2^2 {\tilde g_u'}{}^2 y_b-\frac{1}{150} g_1^4 y_b+\frac{1}{2} g_2^4 y_b-\frac{9}{16} {\tilde g_d'}{}^4 y_b-\frac{5}{4} {\tilde g_d'}{}^2 {\tilde g_u'}{}^2 y_b+\frac{5}{8} {\tilde g_d'}{}^2 y_b y_t^2-\frac{9}{8} {\tilde g_d'}{}^2 y_b^3-\frac{9 {\tilde g_u'}{}^4 y_b}{16}+\frac{5}{8} {\tilde g_u'}{}^2 y_b y_t^2-\frac{9}{8} {\tilde g_u'}{}^2 y_b^3$ \vspace*{0.25cm}

\hspace*{-1.9cm}  $ \Delta b^{(1)}_{y_{\tau }} =\frac{{\tilde g_d'}{}^2 y_{\tau }}{2}+\frac{{\tilde g_u'}{}^2 y_{\tau }}{2}$ \vspace*{0.25cm}

\hspace*{-1.9cm}  $ \Delta b^{(2)}_{y_{\tau }} =\frac{3}{16} g_1^2 {\tilde g_d'}{}^2 y_{\tau }+\frac{15}{16} g_2^2 {\tilde g_d'}{}^2 y_{\tau }+\frac{3}{16} g_1^2 {\tilde g_u'}{}^2 y_{\tau }+\frac{15}{16} g_2^2 {\tilde g_u'}{}^2 y_{\tau }+\frac{33}{50} g_1^4 y_{\tau }+\frac{1}{2} g_2^4 y_{\tau }-\frac{9}{16} {\tilde g_d'}{}^4 y_{\tau }-\frac{5}{4} {\tilde g_d'}{}^2 {\tilde g_u'}{}^2 y_{\tau }-\frac{9}{8} {\tilde g_d'}{}^2 y_{\tau }^3-\frac{9 {\tilde g_u'}{}^4 y_{\tau }}{16}-\frac{9}{8} {\tilde g_u'}{}^2 y_{\tau }^3$ \vspace*{0.25cm}

\hspace*{-1.9cm}  $ \Delta b^{(1)}_{\lambda } =-{\tilde g_d'}{}^4-2 {\tilde g_d'}{}^2 {\tilde g_u'}{}^2+2 {\tilde g_d'}{}^2 \lambda _H-{\tilde g_u'}{}^4+2 {\tilde g_u'}{}^2 \lambda _H$ \vspace*{0.25cm}

\hspace*{-1.9cm}  $ \Delta b^{(2)}_{\lambda } =\frac{3}{4} g_1^2 {\tilde g_d'}{}^2 \lambda _H+\frac{15}{4} g_2^2 {\tilde g_d'}{}^2 \lambda _H-\frac{9}{100} g_1^4 {\tilde g_d'}{}^2-\frac{3}{4} g_2^4 {\tilde g_d'}{}^2-\frac{3}{10} g_1^2 g_2^2 {\tilde g_d'}{}^2+\frac{3}{4} g_1^2 {\tilde g_u'}{}^2 \lambda _H+\frac{15}{4} g_2^2 {\tilde g_u'}{}^2 \lambda _H-\frac{9}{100} g_1^4 {\tilde g_u'}{}^2-\frac{3}{4} g_2^4 {\tilde g_u'}{}^2-\frac{3}{10} g_1^2 g_2^2 {\tilde g_u'}{}^2+\frac{3}{5} g_1^4 \lambda _H+5 g_2^4 \lambda _H-\frac{36 g_1^6}{125}-4 g_2^6-\frac{4}{5} g_1^2 g_2^4-\frac{12}{25} g_1^4 g_2^2+\frac{5 {\tilde g_d'}{}^6}{2}+\frac{17 {\tilde g_d'}{}^4 {\tilde g_u'}{}^2}{2}-\frac{{\tilde g_d'}{}^4 \lambda _H}{4}+\frac{17 {\tilde g_d'}{}^2 {\tilde g_u'}{}^4}{2}+3 {\tilde g_d'}{}^2 {\tilde g_u'}{}^2 \lambda _H-12 {\tilde g_d'}{}^2 \lambda _H^2+\frac{5 {\tilde g_u'}{}^6}{2}-\frac{{\tilde g_u'}{}^4 \lambda _H}{4}-12 {\tilde g_u'}{}^2 \lambda _H^2$ \vspace*{0.25cm}

\hspace*{-1.9cm}  $ b^{(1)}_{{\tilde g_u'}{}} =3 {\tilde g_u'}{} y_b^2-\frac{9 g_1^2 {\tilde g_u'}{}}{20}-\frac{9 g_2^2 {\tilde g_u'}{}}{4}+2 {\tilde g_d'}{}^2 {\tilde g_u'}{}+\frac{5 {\tilde g_u'}{}^3}{4}+3 {\tilde g_u'}{} y_t^2+{\tilde g_u'}{} y_{\tau }^2$ \vspace*{0.25cm}

\hspace*{-1.9cm}  $ b^{(2)}_{{\tilde g_u'}{}} =\frac{5}{8} g_1^2 {\tilde g_u'}{} y_b^2+\frac{45}{8} g_2^2 {\tilde g_u'}{} y_b^2+20 g_3^2 {\tilde g_u'}{} y_b^2-\frac{21}{4} {\tilde g_d'}{}^2 {\tilde g_u'}{} y_b^2-\frac{27}{8} {\tilde g_u'}{}^3 y_b^2+\frac{3}{2} {\tilde g_u'}{} y_b^2 y_t^2-\frac{27 {\tilde g_u'}{} y_b^4}{4}+\frac{3}{40} g_1^2 {\tilde g_d'}{}^2 {\tilde g_u'}{}+\frac{39}{8} g_2^2 {\tilde g_d'}{}^2 {\tilde g_u'}{}+\frac{309}{160} g_1^2 {\tilde g_u'}{}^3+\frac{165}{32} g_2^2 {\tilde g_u'}{}^3+\frac{17}{8} g_1^2 {\tilde g_u'}{} y_t^2+\frac{45}{8} g_2^2 {\tilde g_u'}{} y_t^2+20 g_3^2 {\tilde g_u'}{} y_t^2+\frac{15}{8} g_1^2 {\tilde g_u'}{} y_{\tau }^2+\frac{15}{8} g_2^2 {\tilde g_u'}{} y_{\tau }^2+\frac{117 g_1^4 {\tilde g_u'}{}}{200}-\frac{21 g_2^4 {\tilde g_u'}{}}{4}-\frac{27}{20} g_1^2 g_2^2 {\tilde g_u'}{}-\frac{9 {\tilde g_d'}{}^4 {\tilde g_u'}{}}{4}-\frac{15 {\tilde g_d'}{}^2 {\tilde g_u'}{}^3}{4}-3 {\tilde g_d'}{}^2 {\tilde g_u'}{} \lambda _H-\frac{21}{4} {\tilde g_d'}{}^2 {\tilde g_u'}{} y_t^2-\frac{7}{4} {\tilde g_d'}{}^2 {\tilde g_u'}{} y_{\tau }^2-\frac{3 {\tilde g_u'}{}^5}{4}-3 {\tilde g_u'}{}^3 \lambda _H-\frac{27}{8} {\tilde g_u'}{}^3 y_t^2-\frac{9}{8} {\tilde g_u'}{}^3 y_{\tau }^2+\frac{3 {\tilde g_u'}{} \lambda _H^2}{2}-\frac{27 {\tilde g_u'}{} y_t^4}{4}-\frac{9 {\tilde g_u'}{} y_{\tau }^4}{4}$ \vspace*{0.25cm}

\hspace*{-1.9cm}  $ b^{(1)}_{{\tilde g_d'}{}} =3 {\tilde g_d'}{} y_b^2-\frac{9 g_1^2 {\tilde g_d'}{}}{20}-\frac{9 g_2^2 {\tilde g_d'}{}}{4}+\frac{5 {\tilde g_d'}{}^3}{4}+2 {\tilde g_d'}{} {\tilde g_u'}{}^2+3 {\tilde g_d'}{} y_t^2+{\tilde g_d'}{} y_{\tau }^2$ \vspace*{0.25cm}

\hspace*{-1.9cm}  $ b^{(2)}_{{\tilde g_d'}{}} =\frac{5}{8} g_1^2 {\tilde g_d'}{} y_b^2+\frac{45}{8} g_2^2 {\tilde g_d'}{} y_b^2+20 g_3^2 {\tilde g_d'}{} y_b^2-\frac{27}{8} {\tilde g_d'}{}^3 y_b^2-\frac{21}{4} {\tilde g_d'}{} {\tilde g_u'}{}^2 y_b^2+\frac{3}{2} {\tilde g_d'}{} y_b^2 y_t^2-\frac{27 {\tilde g_d'}{} y_b^4}{4}+\frac{309}{160} g_1^2 {\tilde g_d'}{}^3+\frac{165}{32} g_2^2 {\tilde g_d'}{}^3+\frac{3}{40} g_1^2 {\tilde g_d'}{} {\tilde g_u'}{}^2+\frac{39}{8} g_2^2 {\tilde g_d'}{} {\tilde g_u'}{}^2+\frac{17}{8} g_1^2 {\tilde g_d'}{} y_t^2+\frac{45}{8} g_2^2 {\tilde g_d'}{} y_t^2+20 g_3^2 {\tilde g_d'}{} y_t^2+\frac{15}{8} g_1^2 {\tilde g_d'}{} y_{\tau }^2+\frac{15}{8} g_2^2 {\tilde g_d'}{} y_{\tau }^2+\frac{117 g_1^4 {\tilde g_d'}{}}{200}-\frac{21 g_2^4 {\tilde g_d'}{}}{4}-\frac{27}{20} g_1^2 g_2^2 {\tilde g_d'}{}-\frac{3 {\tilde g_d'}{}^5}{4}-\frac{15 {\tilde g_d'}{}^3 {\tilde g_u'}{}^2}{4}-3 {\tilde g_d'}{}^3 \lambda _H-\frac{27}{8} {\tilde g_d'}{}^3 y_t^2-\frac{9}{8} {\tilde g_d'}{}^3 y_{\tau }^2-\frac{9 {\tilde g_d'}{} {\tilde g_u'}{}^4}{4}-3 {\tilde g_d'}{} {\tilde g_u'}{}^2 \lambda _H-\frac{21}{4} {\tilde g_d'}{} {\tilde g_u'}{}^2 y_t^2-\frac{7}{4} {\tilde g_d'}{} {\tilde g_u'}{}^2 y_{\tau }^2+\frac{3 {\tilde g_d'}{} \lambda _H^2}{2}-\frac{27 {\tilde g_d'}{} y_t^4}{4}-\frac{9 {\tilde g_d'}{} y_{\tau }^4}{4}$ \vspace*{0.25cm}

\end{changemargin}

\subsection{SM + Majorana Triplet Fermion + Dirac Doublet Fermion}

\begin{changemargin}{1.22cm}{0cm}

\hspace*{-1.9cm}  $ \Delta b^{(1)}_{g_1} =\frac{2 g_1^3}{5}$ \vspace*{0.25cm}

\hspace*{-1.9cm}  $ \Delta b^{(2)}_{g_1} =-\frac{9}{20} g_1^3 \tilde{g}_d^2-\frac{9}{20} g_1^3 \tilde{g}_u^2+\frac{9 g_1^5}{50}+\frac{9}{10} g_2^2 g_1^3$ \vspace*{0.25cm}

\hspace*{-1.9cm}  $ \Delta b^{(1)}_{g_2} =2 g_2^3$ \vspace*{0.25cm}

\hspace*{-1.9cm}  $ \Delta b^{(2)}_{g_2} =-\frac{11}{4} g_2^3 \tilde{g}_d^2-\frac{11}{4} g_2^3 \tilde{g}_u^2+\frac{59 g_2^5}{2}+\frac{3}{10} g_1^2 g_2^3$ \vspace*{0.25cm}

\hspace*{-1.9cm}  $ \Delta b^{(1)}_{y_t} =\frac{3}{2} y_t \tilde{g}_d^2+\frac{3}{2} y_t \tilde{g}_u^2$ \vspace*{0.25cm}

\hspace*{-1.9cm}  $ \Delta b^{(2)}_{y_t} =\frac{15}{8} y_b^2 y_t \tilde{g}_d^2+\frac{15}{8} y_b^2 y_t \tilde{g}_u^2-\frac{3}{4} y_t \tilde{g}_d^2 \tilde{g}_u^2+\frac{9}{16} g_1^2 y_t \tilde{g}_d^2-\frac{45}{16} y_t \tilde{g}_d^4-\frac{27}{8} y_t^3 \tilde{g}_d^2+\frac{165}{16} g_2^2 y_t \tilde{g}_d^2+\frac{9}{16} g_1^2 y_t \tilde{g}_u^2-\frac{45}{16} y_t \tilde{g}_u^4-\frac{27}{8} y_t^3 \tilde{g}_u^2+\frac{165}{16} g_2^2 y_t \tilde{g}_u^2+\frac{29}{150} g_1^4 y_t+\frac{3}{2} g_2^4 y_t$ \vspace*{0.25cm}

\hspace*{-1.9cm}  $ \Delta b^{(1)}_{y_b} =\frac{3}{2} y_b \tilde{g}_d^2+\frac{3}{2} y_b \tilde{g}_u^2$ \vspace*{0.25cm}

\hspace*{-1.9cm}  $ \Delta b^{(2)}_{y_b} =\frac{15}{8} y_b y_t^2 \tilde{g}_d^2-\frac{3}{4} y_b \tilde{g}_d^2 \tilde{g}_u^2+\frac{9}{16} g_1^2 y_b \tilde{g}_d^2-\frac{45}{16} y_b \tilde{g}_d^4-\frac{27}{8} y_b^3 \tilde{g}_d^2+\frac{165}{16} g_2^2 y_b \tilde{g}_d^2+\frac{15}{8} y_b y_t^2 \tilde{g}_u^2+\frac{9}{16} g_1^2 y_b \tilde{g}_u^2-\frac{45}{16} y_b \tilde{g}_u^4-\frac{27}{8} y_b^3 \tilde{g}_u^2+\frac{165}{16} g_2^2 y_b \tilde{g}_u^2-\frac{1}{150} g_1^4 y_b+\frac{3}{2} g_2^4 y_b$ \vspace*{0.25cm}

\hspace*{-1.9cm}  $ \Delta b^{(1)}_{y_{\tau }} =\frac{3}{2} y_{\tau } \tilde{g}_d^2+\frac{3}{2} y_{\tau } \tilde{g}_u^2$ \vspace*{0.25cm}

\hspace*{-1.9cm}  $ \Delta b^{(2)}_{y_{\tau }} =-\frac{3}{4} y_{\tau } \tilde{g}_d^2 \tilde{g}_u^2+\frac{9}{16} g_1^2 y_{\tau } \tilde{g}_d^2-\frac{45}{16} y_{\tau } \tilde{g}_d^4-\frac{27}{8} y_{\tau }^3 \tilde{g}_d^2+\frac{165}{16} g_2^2 y_{\tau } \tilde{g}_d^2+\frac{9}{16} g_1^2 y_{\tau } \tilde{g}_u^2-\frac{45}{16} y_{\tau } \tilde{g}_u^4-\frac{27}{8} y_{\tau }^3 \tilde{g}_u^2+\frac{165}{16} g_2^2 y_{\tau } \tilde{g}_u^2+\frac{33}{50} g_1^4 y_{\tau }+\frac{3}{2} g_2^4 y_{\tau }$ \vspace*{0.25cm}

\hspace*{-1.9cm}  $ \Delta b^{(1)}_{\lambda } =6 \lambda _H \tilde{g}_d^2-2 \tilde{g}_d^2 \tilde{g}_u^2-5 \tilde{g}_d^4+6 \lambda _H \tilde{g}_u^2-5 \tilde{g}_u^4$ \vspace*{0.25cm}

\hspace*{-1.9cm}  $ \Delta b^{(2)}_{\lambda } =-11 \lambda _H \tilde{g}_d^2 \tilde{g}_u^2+\frac{9}{4} g_1^2 \lambda _H \tilde{g}_d^2-\frac{5}{4} \lambda _H \tilde{g}_d^4-36 \lambda _H^2 \tilde{g}_d^2+\frac{165}{4} g_2^2 \lambda _H \tilde{g}_d^2+\frac{7}{2} \tilde{g}_d^2 \tilde{g}_u^4+\frac{7}{2} \tilde{g}_d^4 \tilde{g}_u^2-8 g_2^2 \tilde{g}_d^2 \tilde{g}_u^2-\frac{27}{100} g_1^4 \tilde{g}_d^2+\frac{63}{10} g_2^2 g_1^2 \tilde{g}_d^2+\frac{47 \tilde{g}_d^6}{2}-20 g_2^2 \tilde{g}_d^4-\frac{153}{4} g_2^4 \tilde{g}_d^2+\frac{9}{4} g_1^2 \lambda _H \tilde{g}_u^2-\frac{5}{4} \lambda _H \tilde{g}_u^4-36 \lambda _H^2 \tilde{g}_u^2+\frac{165}{4} g_2^2 \lambda _H \tilde{g}_u^2-\frac{27}{100} g_1^4 \tilde{g}_u^2+\frac{63}{10} g_2^2 g_1^2 \tilde{g}_u^2+\frac{47 \tilde{g}_u^6}{2}-20 g_2^2 \tilde{g}_u^4-\frac{153}{4} g_2^4 \tilde{g}_u^2+\frac{3}{5} g_1^4 \lambda _H+15 g_2^4 \lambda _H-\frac{36 g_1^6}{125}-\frac{12}{25} g_2^2 g_1^4-\frac{12}{5} g_2^4 g_1^2-12 g_2^6$ \vspace*{0.25cm}

\hspace*{-1.9cm}  $ b^{(1)}_{\tilde{g}_u} =3 y_b^2 \tilde{g}_u+\tilde{g}_d^2 \tilde{g}_u+3 y_t^2 \tilde{g}_u+y_{\tau }^2 \tilde{g}_u+\frac{11 \tilde{g}_u^3}{4}-\frac{9}{20} g_1^2 \tilde{g}_u-\frac{33}{4} g_2^2 \tilde{g}_u$ \vspace*{0.25cm}

\hspace*{-1.9cm}  $ b^{(2)}_{\tilde{g}_u} =\frac{3}{4} y_b^2 \tilde{g}_d^2 \tilde{g}_u+\frac{3}{2} y_b^2 y_t^2 \tilde{g}_u-\frac{45}{8} y_b^2 \tilde{g}_u^3-\frac{27}{4} y_b^4 \tilde{g}_u+\frac{5}{8} g_1^2 y_b^2 \tilde{g}_u+\frac{45}{8} g_2^2 y_b^2 \tilde{g}_u+20 g_3^2 y_b^2 \tilde{g}_u-\lambda _H \tilde{g}_d^2 \tilde{g}_u+\frac{3}{4} y_t^2 \tilde{g}_d^2 \tilde{g}_u+\frac{1}{4} y_{\tau }^2 \tilde{g}_d^2 \tilde{g}_u-\frac{27}{8} \tilde{g}_d^2 \tilde{g}_u^3-\frac{11}{8} \tilde{g}_d^4 \tilde{g}_u+\frac{3}{20} g_1^2 \tilde{g}_d^2 \tilde{g}_u+\frac{17}{4} g_2^2 \tilde{g}_d^2 \tilde{g}_u-5 \lambda _H \tilde{g}_u^3+\frac{3}{2} \lambda _H^2 \tilde{g}_u-\frac{45}{8} y_t^2 \tilde{g}_u^3-\frac{27}{4} y_t^4 \tilde{g}_u+\frac{17}{8} g_1^2 y_t^2 \tilde{g}_u+\frac{45}{8} g_2^2 y_t^2 \tilde{g}_u+20 g_3^2 y_t^2 \tilde{g}_u-\frac{15}{8} y_{\tau }^2 \tilde{g}_u^3-\frac{9}{4} y_{\tau }^4 \tilde{g}_u+\frac{15}{8} g_1^2 y_{\tau }^2 \tilde{g}_u+\frac{15}{8} g_2^2 y_{\tau }^2 \tilde{g}_u-\frac{7 \tilde{g}_u^5}{2}+\frac{87}{32} g_1^2 \tilde{g}_u^3+\frac{875}{32} g_2^2 \tilde{g}_u^3+\frac{117}{200} g_1^4 \tilde{g}_u-\frac{409}{12} g_2^4 \tilde{g}_u+\frac{9}{20} g_1^2 g_2^2 \tilde{g}_u$ \vspace*{0.25cm}

\hspace*{-1.9cm}  $ b^{(1)}_{\tilde{g}_d} =3 y_b^2 \tilde{g}_d+3 y_t^2 \tilde{g}_d+\tilde{g}_d \tilde{g}_u^2+y_{\tau }^2 \tilde{g}_d+\frac{11 \tilde{g}_d^3}{4}-\frac{9}{20} g_1^2 \tilde{g}_d-\frac{33}{4} g_2^2 \tilde{g}_d$ \vspace*{0.25cm}

\hspace*{-1.9cm}  $ b^{(2)}_{\tilde{g}_d} =\frac{3}{2} y_b^2 y_t^2 \tilde{g}_d+\frac{3}{4} y_b^2 \tilde{g}_d \tilde{g}_u^2-\frac{45}{8} y_b^2 \tilde{g}_d^3-\frac{27}{4} y_b^4 \tilde{g}_d+\frac{5}{8} g_1^2 y_b^2 \tilde{g}_d+\frac{45}{8} g_2^2 y_b^2 \tilde{g}_d+20 g_3^2 y_b^2 \tilde{g}_d-\lambda _H \tilde{g}_d \tilde{g}_u^2-5 \lambda _H \tilde{g}_d^3+\frac{3}{2} \lambda _H^2 \tilde{g}_d+\frac{3}{4} y_t^2 \tilde{g}_d \tilde{g}_u^2-\frac{45}{8} y_t^2 \tilde{g}_d^3-\frac{27}{4} y_t^4 \tilde{g}_d+\frac{17}{8} g_1^2 y_t^2 \tilde{g}_d+\frac{45}{8} g_2^2 y_t^2 \tilde{g}_d+20 g_3^2 y_t^2 \tilde{g}_d+\frac{1}{4} y_{\tau }^2 \tilde{g}_d \tilde{g}_u^2-\frac{27}{8} \tilde{g}_d^3 \tilde{g}_u^2-\frac{11}{8} \tilde{g}_d \tilde{g}_u^4+\frac{3}{20} g_1^2 \tilde{g}_d \tilde{g}_u^2+\frac{17}{4} g_2^2 \tilde{g}_d \tilde{g}_u^2-\frac{15}{8} y_{\tau }^2 \tilde{g}_d^3-\frac{9}{4} y_{\tau }^4 \tilde{g}_d+\frac{15}{8} g_1^2 y_{\tau }^2 \tilde{g}_d+\frac{15}{8} g_2^2 y_{\tau }^2 \tilde{g}_d-\frac{7 \tilde{g}_d^5}{2}+\frac{87}{32} g_1^2 \tilde{g}_d^3+\frac{875}{32} g_2^2 \tilde{g}_d^3+\frac{117}{200} g_1^4 \tilde{g}_d-\frac{409}{12} g_2^4 \tilde{g}_d+\frac{9}{20} g_1^2 g_2^2 \tilde{g}_d$ \vspace*{0.25cm}

\end{changemargin}

\section{One-Loop Threshold Corrections}
\label{app:thresholds}

For a fully self-consistent next-to-leading order calculation, one has to include one-loop threshold effects at the weak scale.  The pole Higgs and top masses are related to the quartic $\lambda_H$ and top Yukawa coupling $y_t$ via
\bea
m_H^2 \left(1+\delta_H\right) &=& 2  \lambda_H v^2\\
m_t (1+ \delta_t) &=& y_t v,
\eea
where the correction $\delta_H$ is equal to
\bea
\delta_H &=& \frac{G_F m_Z^2}{8 \sqrt{2} \pi ^2} \left(x_H F_1+ F_0 + x_H^{-1} F_{-1}  + F_{NP}\right ) \\
\delta_t &=& \delta_{t, SM} + \delta_{t,NP}
   \eea
Here the functions $F_1$, $F_0$, $F_{-1}$ are SM contributions to the Higgs quartic which can be found in~\cite{Sirlin}, while $\delta_{t,SM}$ are SM contributions to the top Yukawa which can be found in~\cite{Giudice:2011cg}.

The quantities $F_{NP}$ and $G_{NP}$ denotes new physics contributions.
    Corrections for the singlet/doublet or doublet/triplet fermions may be found in~\cite{Giudice:2011cg}, who use a slightly different notation.  We have calculated the one-loop thresholds for the scalar models utilizing the methods of~\cite{Sirlin,Pierce}. The corrections to the Higgs quartic from a singlet, doublet and triplet scalars are

   \bea
F_S &=&
   \frac{2 \left(\xi _S-\xi
   _{S_0}\right){}^2 \left(-2 \sqrt{\frac{4 \xi
   _{S_0}}{\xi _H}-1} \cot ^{-1}\left(\sqrt{\frac{4 \xi
   _{S_0}}{\xi _H}-1}\right)+\log \left(\frac{Q^2}{m_Z^2
   \xi _{S_0}}\right)+2\right)}{\xi _H} \nonumber\\
F_T &=& 
\frac{6 \left(\xi _T-\xi
   _{T_0}\right){}^2 \left(-2 \sqrt{\frac{4 \xi
   _{T_0}}{\xi _H}-1} \cot ^{-1}\left(\sqrt{\frac{4 \xi
   _{T_0}}{\xi _H}-1}\right)+\log \left(\frac{Q^2}{m_Z^2 \xi
   _{T_0}}\right)+2\right)}{\xi _H} \nonumber\\ 
F_D &=&
\frac{4 \left(2 \xi _D^2-2
   \left(\xi _{D_+}+\xi _{D_0}\right) \xi _D+\xi
   _{D_+}^2+\xi _{D_0}^2\right) \log
   \left(\frac{Q^2}{m_Z^2}\right)}{\xi
   _H} \nonumber \\
   &&+\frac{\left(\xi _{D_+}+\xi _{D_0}\right) \xi _H+16
   \xi _D^2-16 \left(\xi _{D_+}+\xi _{D_0}\right) \xi
   _D+8 \left(\xi _{D_+}^2+\xi _{D_0}^2\right)}{\xi
   _H} \nonumber\\
   &&+\left(\frac{2 \xi _{D_+} \xi _{D_0}}{\xi
   _{D_+}-\xi _{D_0}}-\frac{4 \left(\xi _D-\xi
   _{D_0}\right){}^2}{\xi _H}\right) \log \left(\xi
   _{D_0}\right)+\left(\frac{2 \xi _{D_+} \xi _{D_0}}{\xi
   _{D_0}-\xi _{D_+}}-\frac{4 \left(\xi _D-\xi
   _{D_+}\right){}^2}{\xi _H}\right) \log \left(\xi
   _{D_+}\right)\nonumber\\
   &&-\frac{8 \left(\xi _D-\xi
   _{D_+}\right){}^2 \sqrt{\frac{4 \xi _{D_+}}{\xi _H}-1}
   \cot ^{-1}\left(\sqrt{\frac{4 \xi _{D_+}}{\xi
   _H}-1}\right)}{\xi _H}-\frac{8 \left(\xi _D-\xi
   _{D_0}\right){}^2 \sqrt{\frac{4 \xi _{D_0}}{\xi _H}-1}
   \cot ^{-1}\left(\sqrt{\frac{4 \xi _{D_0}}{\xi
   _H}-1}\right)}{\xi _H}.\nonumber\\
\eea
Here we have defined $\xi_X = m_X^2 / m_Z^2$.  Note that while $m_{S,T,D}$ denote the Lagrangian parameters defined in \Sec{sec:models}, we have also defined that
\bea
m_{S_0}^2 &=&  m_S^2 +  \kappa_S v^2 \\
m_{T_0}^2 &=&  m_T^2 + \kappa_T v^2 \\
m_{D_0}^2 &=&  m_D^2 + \kappa_D v^2/ 2 \\
m_{D_+}^2 &=&  m_D^2 + (\kappa_D+\kappa_D') v^2/ 2,
\eea
which denote the physical masses within the singlet, triplet, and doublet multiplets.

Finally, we present the threshold correction to the top quark Yukawa coupling from new physics, which reads
\bea
\delta_{t,D}&=&\frac{g_2^2 \left(-x_{D_+}^2+x_{D_0}^2+2 x_{D_0} x_{D_+}
   \log \left(\frac{x_{D_+}}{x_{D_0}}\right)\right) \sec
   ^2\left(\theta _W\right)}{128 \pi ^2
   \left(x_{D_0}-x_{D_+}\right)}
\eea
for the scalar doublet and vanishes in the case of a scalar singlet and triplet.

\end{document}